\documentclass[twocolumn, apj, twocolappendix]{aastex631}

\usepackage{lipsum}
\usepackage{amsmath}
\usepackage{enumerate}

\begin{document}

\shorttitle{TOI-2533 $b$'s Spin-Orbit Alignment}
\shortauthors{Ferreira et al.}

\title{SOLES XII. \\ The Aligned Orbit of TOI-2533 b, a Transiting Brown Dwarf Orbiting an F8-type Star}

\correspondingauthor{Thiago Ferreira dos Santos}
\email{thiago.dossantos@yale.edu}

\author[0000-0003-2059-470X]{Thiago Ferreira}
\affiliation{Department of Astronomy, Yale University, 219 Prospect Street, New Haven, CT 06511, USA}

\author[0000-0002-7670-670X]{Malena Rice}
\affiliation{Department of Astronomy, Yale University, 219 Prospect Street, New Haven, CT 06511, USA}

\author[0000-0002-0376-6365]{Xian-Yu Wang}
\affiliation{Department of Astronomy, Indiana University, 727 East 3rd Street, Bloomington, IN 47405-7105, USA}

\author[0000-0002-7846-6981]{Songhu Wang}
\affiliation{Department of Astronomy, Indiana University, 727 East 3rd Street, Bloomington, IN 47405-7105, USA}

\begin{abstract}

Brown dwarfs occupy a middle ground in mass space between gaseous giant planets and ultra-cool dwarf stars, and the characterisation of their orbital orientations may shed light on how these neighbouring objects form. We present an analysis of the Rossiter-McLaughlin (RM) effect across the transit of TOI-2533 $b$, a brown dwarf on a moderately eccentric ($e_b = 0.2476\pm0.0090$) and wide-separation ($a_b/R_\star = 13.34\pm0.30$) orbit around an F8-type star, using data from the NEID/WIYN spectrograph in combination with archival photometry and radial velocity observations. Spin-orbit analyses of brown dwarfs are relatively rare, and TOI-2533 stands out as the fifth brown dwarf system with a measured spin-orbit constraint. We derive a sky-projected stellar obliquity of $\lambda = -7\pm14^{\circ}$ for TOI-2533 $b$, finding that the brown dwarf is consistent with spin-orbit alignment. Our joint model also indicates that TOI-2533 $b$ falls near the lower bound of the hydrogen-burning minimum mass range (M$_b$ = $74.9\pm5.3$ M$_{\rm \tiny Jup}$). Ultimately, we find that TOI-2533 $b$ is consistent with formation from disc fragmentation in a primordially spin-orbit aligned orientation, although we cannot rule out the possibility that the system has been tidally realigned during its lifetime. 
    
\end{abstract}

\keywords{Brown dwarfs (185) --- Exoplanet dynamics (490) --- Exoplanet formation (492) --- Exoplanet evolution (491) --- Planetary alignment (1243) --- Star-planet interactions (2177) --- Binary stars (154) --- Close binary stars (254)}

\section{Introduction}\label{sec:introduction}

The formation of eclipsing systems -- including exoplanets, brown dwarfs, and stars -- can be constrained at a population level, and in some cases at an individual-system level, through precise measurements of the masses and relative orbits of the system components \citep[e.g.][]{2009ApJ...703.2091W, 2010ApJ...718L.145W, 2010ApJ...719..602S, 2012ApJ...757...18A, 2013ApJ...767...32A, 2022AJ....164..104R, 2024AJ....167..126R}. The key formation channels of brown dwarfs, whether star-like or planet-like (or a mix), have been long contested \citep[e.g.][]{2002MNRAS.332L..65B, 2007A&A...466..943G, 2008MNRAS.389.1556B, 2017A&A...602A..38M, 2020AJ....159...63B}. While a rough mass cutoff is generally used for convenience to delineate the borders between planets ($M\lesssim13M_{\rm \tiny Jup}$), brown dwarfs ($M \sim 13-80M_{\rm \tiny Jup}$), and stars ($M\gtrsim80M_{\rm \tiny Jup}$), an improved understanding of the formation mechanisms that separate each regime, including what properties typically characterise each group, is needed to define physically motivated boundaries (e.g., \citealt{2018ApJ...853...37S}).

Divergent eccentricity distributions for sub-stellar companions suggest distinct formation mechanisms at the high- and low-mass end. For example, \citet{2014MNRAS.439.2781M} showed that systems with low-mass ($M_{\rm BD} < 42.5$ M$_{\rm \tiny Jup}$) brown dwarf companions demonstrate an anticorrelation between mass and maximum eccentricity, whereas this trend is not apparent for systems with high-mass brown dwarf companions ($M_{\rm BD} > 42.5$ M$_{\rm \tiny Jup}$), which follow an eccentricity distribution more similar to that of stellar binaries. \citet{2020AJ....159...63B} more recently showed that high-mass-ratio systems ($M_{\rm BD}/M_*>0.01$) follow a broader eccentricity distribution than low-mass-ratio systems ($M_{\rm BD}/M_*<0.01$), which tend toward lower eccentricities. These findings suggest that low-mass ($M_{\rm BD} < 42.5$ M$_{\rm \tiny Jup}$) and low-mass-ratio ($q<0.01$) brown dwarf companions may form through gravitational instability within circumstellar discs \citep{1997Sci...276.1836B}, while high-mass ($M_{\rm BD} > 42.5~M_{\rm \tiny Jup}$) and high-mass-ratio ($M_{\rm BD}/M_*>0.01$) brown dwarf companions instead form more like stars, through molecular cloud collapse \citep{1969MNRAS.145..271L, 2002ARA&A..40..349T}.

Improved characterisation of brown dwarf systems, particularly in light of a growing population of known transiting brown dwarfs \citep{2020AJ....159..151S, 2021AJ....161...97C, 2022MNRAS.514.4944C, 2022A&A...664A..94P, 2023MNRAS.519.5177C} with the advent of the Transiting Exoplanet Survey Satellite \citep[TESS;][]{2015JATIS...1a4003R}, offers a compelling path forward to further distinguish between planet-like and star-like formation histories. Each formation mechanism should have a distinct imprint on the final distribution of brown dwarf spin-orbit orientations -- that is, the alignment of the brown dwarf's orbit normal with the spin axis of its host star. Turbulent fragmentation, for instance, may result in a diverse range of orientations \citep{2016ApJ...827L..11O}, while disc-based pathways are typically linked to primordial alignment (although see \citet{2003MNRAS.339..577B, 2010MNRAS.401.1505B} for exceptions). 

Spin-orbit orientations can be observationally constrained for transiting systems (see \citet{2022PASP..134h2001A} for a review of this topic). During the transit of a spin-orbit aligned occulter, the blue-shifted hemisphere of a star is first obscured, followed by an equal-duration obscuration of the red-shifted hemisphere \citep{1893AstAp..12..646H}. The corresponding distortion in the net Doppler shift measured across the integrated light from the star during the transit event is known as the Rossiter-McLaughlin (RM) effect \citep{1924ApJ....60...15R, 1924ApJ....60...22M}. 

The RM effect can be measured to derive the sky-projected spin-orbit angle $\lambda$ (a 2D proxy for the true, 3D spin-orbit angle $\psi$) between the stellar spin axis and the occulter's orbit normal. Asymmetries in the RM profile are indicative of orbits that are misaligned with the host star's equator, hinting that a particular system either did not emerge from a spin-orbit aligned circumstellar disc or that it was dynamically perturbed after disc dispersal \citep[e.g.][]{2011Natur.473..187N, 2012ApJ...754L..36N, 2019MNRAS.486.2265T}.

To date\footnote{As of May 2024; based on the Encyclopaedia of Exoplanetary Systems (\url{https://exoplanet.eu/home/}) and the Transiting Exoplanets Catalogue (TEPCat \citep{2011MNRAS.417.2166S}; \url{https://astro.keele.ac.uk/jkt/tepcat/tepcat.html})}, orbital obliquity measurements have been reported for 211 transiting exoplanets and brown dwarfs, with most of the observations focused on short-period, Jovian-mass exoplanet companions. Among these systems, only four RM observations have been reported for transiting brown dwarfs: CoRoT-3b ($\lambda = -37.6^{+10.0}_{-22.3}$ deg, $M = 21.23^{+0.82}_{-0.59}~M_{\rm \tiny Jup}$; \citealt{2009A&A...506..377T}), KELT-1b ($\lambda = 2\pm16$ deg, $M = 27.38\pm0.93~M_{\rm \tiny Jup}$; \citealt{2012ApJ...761..123S}, WASP-30b ($\lambda = 7^{+19}_{-27}$ deg, $M = 62.5\pm1.2~M_{\rm \tiny Jup}$; \citealt{2013A&A...549A..18T}), and HATS-70b ($\lambda = 8.9^{+5.6}_{-4.5}$ deg, $M = 12.9^{+1.8}_{-1.6}~M_{\rm \tiny Jup}$; \citealt{2019AJ....157...31Z}). Brown dwarf spin-orbit observations remain rare, despite their utility to constrain the formation pathways of these systems.

In this paper, we present an RM analysis of TOI-2533 $b$, a brown dwarf with orbital period $P\sim6.68$ days first confirmed by \cite{2023AJ....166..225S} that transits an F8-type star bordering the Kraft break \citep{1967ApJ...150..551K}. Our study leverages observations obtained using the NEID Spectrograph \citep{2016SPIE.9908E..7HS} on the WIYN 3.5-meter Telescope. This is the twelfth result in the Stellar Obliquities in Long-period Exoplanet Systems (SOLES) survey \citep{2021AJ....162..182R, 2022ApJ...926L...8W, 2022AJ....164..104R, 2023AJ....165...65R, 2023ApJ...949L..35H, 2023ApJ...951L..29D, 2023AJ....166..217W, 2023AJ....166..266R, 2023ApJ...959L...5L, 2024AJ....167..175H, 2024arXiv240406504R}, which is designed to extend the set of spin-orbit constraints to relatively wide-orbiting sub-stellar companions. The observations of TOI-2533 obtained within this work are described in Section \ref{sec:observations}, and a description of the stellar fundamental parameters are presented in Section \ref{sec:stellar}. The joint transit, radial velocity, and RM effect models are presented in Section \ref{sec:spin-orbit_modelling}, and we discuss dynamical timescales in Section \ref{sec:dyntimescales}. Insights into the formation and evolution of TOI-2533 $b$ and its placement among other transiting brown dwarfs with measured spin-orbit orientations are presented in Section \ref{sec:discussion}. Our conclusions are then given in Section \ref{sec:conclusions}.

\section{Observations}\label{sec:observations}

We obtained 21 in-transit radial velocity measurements of TOI-2533 (TYC 2010-124-1; Gaia DR3 1259922196651254272; TIC 418012030) in high-resolution mode ($R~\sim~110~000$) with the NEID Spectrograph mounted on the 3.5--meter WIYN Telescope at Kitt Peak National Observatory in Arizona/US \citep{2016SPIE.9908E..7HS}. Observations were performed on April 15, 2023, from 06:37 to 11:42 UT, with exposure times of 900 seconds resulting in an average SNR = 20.8 px$^{-1}$ at $5500$ \AA. Air masses varied from 1.06 at the beginning of the observations to 1.47 at the end, and the collected radial velocities are provided in Table \ref{tab:NEIDData}. The data was reduced using the NEID Data Reduction Pipeline \citep{2019ASPC..523..567K}. 

\begin{table}[t]
    \centering
	\caption{In-transit radial velocities for the TOI-2533 system collected with NEID/WIYN spectrograph.}
    \begin{tabular}{lcc}
    \hline 
	\hline
    Time (BJD) & RV (km s$^{-1}$) & $\sigma_{\rm RV}$ (km s$^{-1}$)\\[0.1cm]
    \hline
    2460049.7864386 & -9.6693 & 0.0114 \\
    2460049.7966859 & -9.7940 & 0.0108 \\
    2460049.8075163 & -9.9099 & 0.0116 \\
    2460049.8182185 & -10.0251 & 0.0122 \\
    2460049.8288248 & -10.1373 & 0.0119 \\
    2460049.8397229 & -10.2756 & 0.0114 \\
    2460049.8498363 & -10.3747 & 0.0115 \\
    2460049.8607308 & -10.4801 & 0.0110 \\
    2460049.8713124 & -10.5995 & 0.0106 \\
    2460049.8821565 & -10.7191 & 0.0114 \\
    2460049.8924628 & -10.8390 & 0.0105 \\
    2460049.9031508 & -10.9848 & 0.0101 \\
    2460049.9136643 & -11.1078 & 0.0101 \\
    2460049.9242635 & -11.2587 & 0.0102 \\
    2460049.9347213 & -11.3692 & 0.0108 \\
    2460049.9454090 & -11.5017 & 0.0107 \\
    2460049.9560650 & -11.6314 & 0.0119 \\
    2460049.9666020 & -11.7174 & 0.0129 \\
    2460049.9773795 & -11.8458 & 0.0136 \\
    2460049.9879091 & -11.9614 & 0.0133 \\
    2460049.9987549 & -12.0692 & 0.0138 \\
    \hline 
    \hline
    \end{tabular}
    \label{tab:NEIDData}
\end{table}

\section{Stellar Characterisation}\label{sec:stellar}

\subsection{Atmospheric Parameters}
We performed a spectral fitting analysis on our co-added out-of-transit NEID spectra, employing the \texttt{iSpec} Python package \citep{Blanco2014, Blanco2019}, to derive stellar atmospheric parameters for TOI-2533. From this fit, we obtained $T_{\rm eff}=6228\pm114$ K, $\log g = 4.45\pm0.18$ cm s$^{-2}$, [Fe/H] = $-0.33\pm0.07$, and $v\sin i_*=5.55\pm0.75$ km/s. Our derived values agree with those from \citet{2023AJ....166..225S} within $2\sigma$.

\subsection{SED Analysis}
We also leveraged the empirical stellar spectra library of \cite{2017ApJS..230...16K} to further characterise the TOI-2533 host star. Our SED fit drew archival data from the Two Micron All Sky Survey (2MASS; \citealt{2006AJ....131.1163S}), \emph{Gaia} DR3 \citep{2023A&A...674A...1G}, the Hubble Space Telescope (HST; \citealt{2011ApJS..193...27W}), the Javalambre Physics of the Accelerating Universe Astrophysical Survey (J-PAS; \citealt{2014arXiv1403.5237B}), the Javalambre-Photometric Local Universe Survey (J-PLUS; \citealt{2019A&A...622A.176C}), the Sloan Digital Sky Survey (SDSS; \citealt{2000AJ....120.1579Y}, the Tycho-2 Catalogue \citep{2000A&A...357..367H}, and the Wide-field Infrared Survey Explorer (WISE; \citealt{2010AJ....140.1868W}).

By integrating the observed Spectral Energy Distribution (SED) from the optical ($\lambda_{\rm min} = 3551.05$ \AA) to the infrared region ($\lambda_{\rm max} = 115608$ \AA) within the Virtual Observatory SED Analyser (VOSA\footnote{\url{http://svo2.cab.inta-csic.es/theory/vosa/}}; \citealt{2008A&A...492..277B}), we found that TOI-2533 is an F8-type star. The mass ($1.10^{+0.01}_{-0.04}~M_\odot$), radius ($1.12\pm0.02~R_\odot$), and effective temperature ($6183^{+16}_{-84}~K$) derived from the SED fitting also agree with the values reported in \cite{2023AJ....166..225S}.

We note that this analysis was conducted independently of the joint fit described in Section \ref{sec:spin-orbit_modelling}. Parameters from \cite{2023AJ....166..225S}, obtained through spectroscopic analyses, were adopted as priors to derive the stellar host's physical and orbital parameters. 

\subsection{Additional System Properties}

We searched for signatures of rotational variability within our TESS photometry of the TOI-2533 star using both the Generalised Lomb-Scargle periodogram (GLS; \citealt{2009A&A...496..577Z}) and an Autocorrelation Function analysis (ACF; \citealt{2009CoAst.160...74H, 2019ApJ...879...33V}). No consistent, statistically significant rotational signal was detected, such that we were unable to derive the degree of line-of-sight alignment or the 3D spin-orbit orientation. We also found no evidence for bound stellar companions to the TOI-2533 system through a search leveraging both \textit{Gaia} DR3 and existing adaptive optics constraints (see Appendix \ref{app:noboundcompanion}).


\section{Joint Spin-Orbit Modelling}\label{sec:spin-orbit_modelling}

\begin{figure*}[t]
    \centering
    \includegraphics[width = \linewidth]{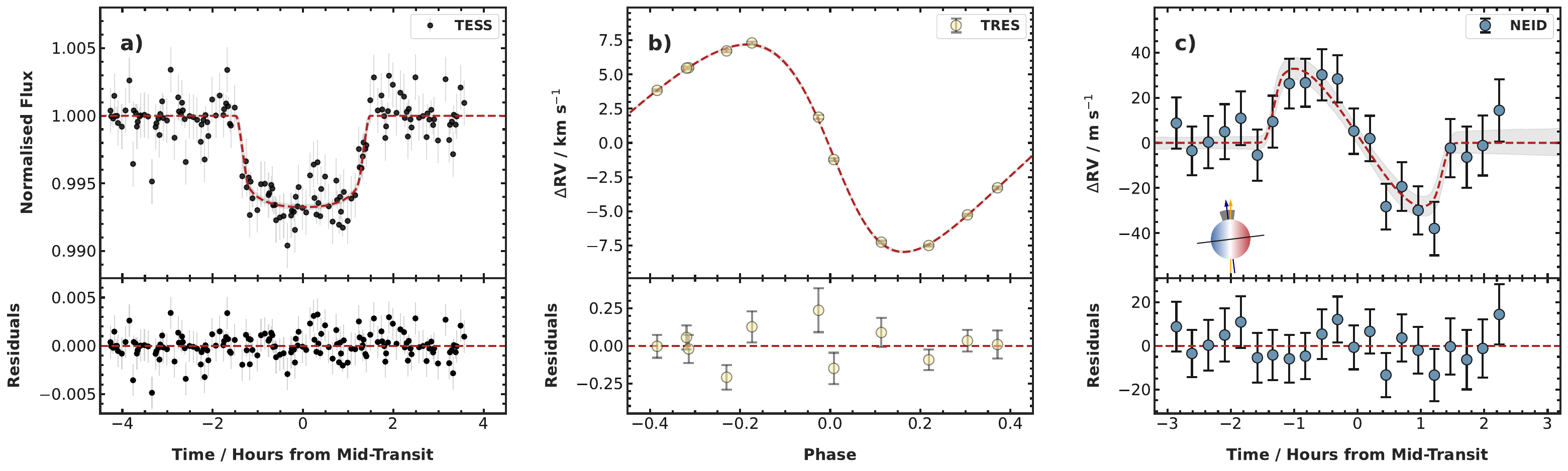}
    \caption{Best-fit joint model for TOI-2533 $b$. Panel (a) presents the phase-folded TESS light curve, panel (b) the TRES radial velocity observations, and panel (c) the NEID Rossiter-McLaughlin observations. The residuals are presented at the bottom of each panel, along with the $1\sigma$ uncertainty of the joint model in grey. A schematic representation of the TOI-2533 $b$ side-view orbit is presented at the bottom left of the panel (c), corresponding to a sky-projected obliquity of $\lambda = -7\pm14^{\circ}$.}
    \label{fig:bestfit_plots}
\end{figure*}

We used the {\sc allesfitter} Python package \citep{2021ApJS..254...13G} to jointly model the new RM measurement along with other publicly available data sets for TOI-2533. The star was observed during TESS Sectors 23 and 50 at $1800$ s and $600$ s cadence, respectively. Sector 23 data were processed by the Science Processing Operations Center (SPOC; \citealt{2016SPIE.9913E..3EJ}) standard aperture pipeline and retrieved using the {\sc lightkurve} software package \citep{2018ascl.soft12013L}, while Sector 50 data were processed using the TESS Quick-Look Pipeline (QLP; \citealt{2022RNAAS...6..236K}) and retrieved from the Mikulski Archive for Space Telescopes (MAST)\footnote{\url{https://exo.mast.stsci.edu}}. Outliers in the data were removed with an upper sigma clipping of 2 away from the median. 

Archival radial velocity data, obtained with the Tillinghast Reflector {\'E}chelle Spectrograph (TRES) spectrograph \citep{2011ASPC..442..305M}, were drawn from \cite{2023AJ....166..225S}. To remove systematic instrumental trends characterised by a gradual increase/decrease in the measured stellar flux over time in the TESS data, we applied a de-trending model following the method outlined in \citep{2022ApJS..259...62I}, where a polynomial of either 1st, 2nd, or 3rd, order is selected based on a minimisation of the Bayesian Information Criterion of the data. 

The free parameters in our joint model include the companion-to-star radius ratio ($R_b / R_\star$), sum of radii divided by the orbital semi-major axis ($(R_\star + R_b) / a_b$), cosine of the orbital inclination ($\cos{i_b}$), transit epoch ($T_{0;b}$), orbital period ($P_b$; with initial guess as reported in \citealt{2023AJ....166..225S}), radial velocity semi-amplitude ($K_b$), orbital eccentricity and argument of periapsis coupled as $\sqrt{e}\cos(\omega)$ and $\sqrt{e}\sin(\omega)$, limb-darkening coefficients in quadratic form for TESS ($q_{1; \mathrm{TESS}}$ and $q_{2; \mathrm{TESS}}$) and NEID ($q_{1; \mathrm{NEID}}$ and $q_{2; \mathrm{NEID}}$) data (see \cite{2013MNRAS.435.2152K} and \cite{2016MNRAS.457.3573E} for details), jitter terms for the NEID and TRES datasets that were added in quadrature, and a white noise flux error scaling term for the TESS photometry (see \citealt{2021ApJS..254...13G}). The sky-projected spin-orbit angle $\lambda$ was allowed to freely vary between $\pm180^\circ$, and the projected stellar rotation velocity $v\sin{i}_\star$ was also set as a free parameter. The priors for each hyper-parameter are summarised in Table \ref{tab:allesfitter_results}, and the host star priors were drawn from \cite{2023AJ....166..225S} -- specifically, $R_\star = 1.11\pm0.01~R_\odot$, $M_\star = 1.02^{+0.06}_{-0.07}~M_\odot$, and $T_{\rm eff, \star} = 6180\pm60$ K.

The Markov hyper-parameter space was probed using the Dynamic Nested Sampling algorithm {\sc dynesty} \citep{2020MNRAS.493.3132S} over 1000 live-points and tolerance of the convergence criterion of $0.01$ (see details in \citealt{2021ApJS..254...13G}). The resulting orbital and physical parameters derived for TOI-2533 $b$ are listed in Table \ref{tab:allesfitter_results}, and the RM best-fit model, along with the transit and RV models, is presented in Figure \ref{fig:bestfit_plots}.

\begin{figure*}[t]
    \centering
    \includegraphics[width = \linewidth]{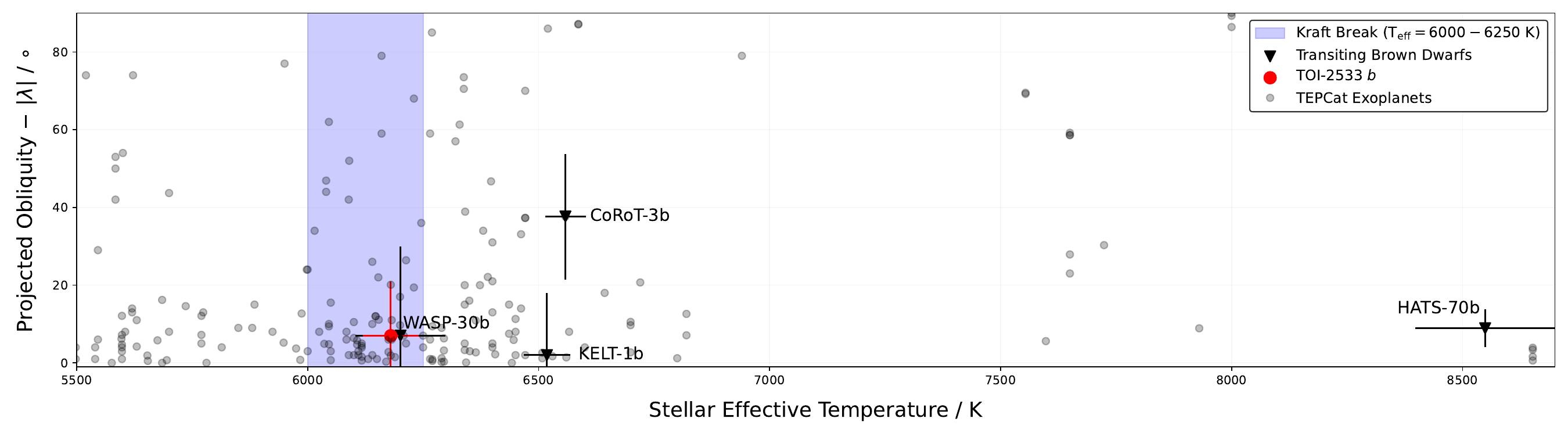}
    \caption{Distribution of the sky-projected spin-orbit angle $\vert\lambda\vert$ as a function of the stellar effective temperature for brown dwarfs and exoplanets. TOI-2533 $b$, reported in this paper, is indicated in red. Background grey dots represent exoplanets presented in the TEPCat catalogue \citep{2011MNRAS.417.2166S} as of May 2024, and black triangles indicate the four known transiting brown dwarfs with previously reported spin-orbit measurements (WASP-30 $b$, CoRoT-3 $b$, KELT-1 $b$, and HATS-70 $b$). The shaded blue area represents the approximate limits of the Kraft break, for which the precise location is a function of the stellar metallicity \citep{2022ApJ...927...22S}.} 
    \label{fig:TEPCat}
\end{figure*}

We find that the TOI-2533 b brown dwarf is consistent with spin-orbit alignment, with $\lambda = -7\pm14$ degrees. This measurement is shown in context with the current census of exoplanet and brown dwarf spin-orbit constraints in Figure \ref{fig:TEPCat}. All other derived system parameters -- corresponding to a brown dwarf with a radius $R_b=0.841\pm0.018~R_{\rm \tiny Jup}$ and mass $74.9\pm5.3~M_{\rm \tiny Jup}$, on a short-period ($P=6.685784\pm0.000013$ days) and moderately eccentric ($e = 0.2476 \pm 0.0090$) orbit -- match with those of \citet{2023AJ....166..225S} within $1\sigma$.

\begin{table*}[t]
    \centering
	\caption{Orbital and Physical Properties derived for TOI-2533 $b$ using \texttt{allesfitter}.\\ {\it Notes}: (a) $\mathcal{U}(\xi\vert\alpha,\beta)$ denotes a uniform distribution between $\alpha$ and $\beta$ with initial guess $\xi$, and $\mathcal{T}(\xi\vert\mu, \sigma, \zeta, \eta)$ denotes a normal distribution with mean $\mu$ and standard deviation $\sigma$, truncated at lower and upper bounds ($\zeta$, $\eta$) with initial guess $\xi$. (b) The transformation between quadratic to linear limb-darkening coefficients is given by $u_1 = 2\sqrt{q_1}q_2$ and $u_2 = \sqrt{q_1}(1-2q_2)$, following \cite{2013MNRAS.435.2152K} and \cite{2021ApJS..254...13G}.}
    \begin{tabular}{llll}
    \hline 
	\hline
    {\bf Description} & {\bf Parameter} & {\bf Value} & {\bf Prior} \\[0.05cm]
    \hline
    \multicolumn{3}{l}{{\bf Fitted Parameters}} \\[0.05cm]
    Orbital period (days)\dotfill & $P_b$ & $6.685784\pm0.000013$ & $\mathcal{U}(6.68577\vert 6.67577,6.69577)$ \\   
    Transit epoch (BJD)\dotfill & $T_{0;b}$ & $2459521.72862_{-0.0010}^{+0.00097}$ & $\mathcal{U}(2459494\vert 2459466,2459522)$ \\   
    Radial velocity semi-amplitude (km s$^{-1}$)\dotfill & $K_b$ & $7.579\pm0.070$ & $\mathcal{U}(7\vert 0,15)$ \\   
    1st eccentricity parameter\dotfill & $\sqrt{e_b} \cos{\omega_b}$ & $-0.104\pm0.013$ & $\mathcal{U}(0\vert -1, 1)$ \\  
    2nd eccentricity parameter\dotfill & $\sqrt{e_b} \sin{\omega_b}$ & $0.487\pm0.010$ & $\mathcal{U}(0\vert -1, 1)$ \\  
    Companion-to-star radius ratio\dotfill & $R_b / R_\star$ & $0.0779\pm0.0015$ & $\mathcal{T}(0.075\vert 0.08,0.4,0,1)$ \\  
    Summed radii divided by semi-major axis\dotfill & $(R_\star + R_b) / a_b$ & $0.0808\pm0.0019$ & $\mathcal{T}(0.077\vert 0.077,0.5,0,1)$ \\  
    Cosine of the orbital inclination\dotfill & $\cos{i_b}$ & $0.0421\pm0.0060$ & $\mathcal{T}(0.0347\vert 0.0685,0.0113,0,1)$ \\  
    Sky-projected spin-orbit angle (deg)\dotfill & $\lambda$ & $-7\pm14$ & $\mathcal{U}(0\vert -180, 180)$ \\
    Stellar rotation velocity (km s$^{-1}$)\dotfill & $v\sin{i}_{*}$ & $8.4_{-1.4}^{+1.5}$ & $\mathcal{U}(5\vert 1, 50)$ \\[0.1cm]
    Quadratic limb-darkening coefficient for TESS\dotfill & $q_{1; \mathrm{TESS}}$ & $0.26_{-0.15}^{+0.23}$ & $\mathcal{U}(0.5\vert 0,1)$ \\  
    Quadratic limb-darkening coefficient for TESS\dotfill & $q_{2; \mathrm{TESS}}$ & $0.19_{-0.14}^{+0.30}$ & $\mathcal{U}(0.5\vert 0,1)$ \\  
    Quadratic limb-darkening coefficient for NEID\dotfill & $q_{1; \mathrm{NEID}}$ & $0.63_{-0.32}^{+0.25}$ & $\mathcal{U}(0.5\vert 0,1)$ \\  
    Quadratic limb-darkening coefficient for NEID\dotfill & $q_{2; \mathrm{NEID}}$ & $0.60_{-0.37}^{+0.28}$ & $\mathcal{U}(0.5\vert 0,1)$ \\  
    TESS jitter\dotfill & $\ln{\sigma_\mathrm{TESS}}$ & $-6.681_{-0.056}^{+0.060}$ & $\mathcal{U}-7.0\vert -15, 0)$ \\
    NEID jitter (km s$^{-1}$)\dotfill & $\ln{\sigma_\mathrm{jit.; NEID}}$ & $-10.6_{-3.0}^{+3.3}$ & $\mathcal{U}(-4.5\vert -15, 0)$ \\
    TRES jitter (km s$^{-1}$)\dotfill & $\ln{\sigma_\mathrm{jit.; TRES}}$ & $-2.38_{-0.67}^{+0.51}$ & $\mathcal{U}(-3.0\vert -15, 0)$ \\
    \hline
    \multicolumn{3}{l}{{\bf Derived Brown Dwarf Parameters}} \\[0.05cm]
    Radius, units of Earth radii ($\mathrm{R_{\oplus}}$)\dotfill & $R_\mathrm{b}$ & $9.43\pm0.20$ & $-$ \\ 
    Radius, units of Jupiter radii ($\mathrm{R_{Jup}}$)\dotfill & $R_\mathrm{b}$ & $0.841\pm0.018$ & $-$ \\ 
    Mass, units of Jupiter masses ($\mathrm{M_{Jup}}$)\dotfill & $M_\mathrm{b}$ & $74.9\pm5.3$ & $-$ \\ 
    Mass, units of solar masses ($\mathrm{M_{\odot}}$)\dotfill & $M_\mathrm{b}$ & $0.0715\pm0.0051$ & $-$ \\ 
    Mass ratio ($M_{b}/M_\star$)\dotfill & $\equiv q_\mathrm{b}$ & $0.0702_{-0.0019}^{+0.0021}$ & $-$ \\ 
    Eccentricity \dotfill & $e_\mathrm{b}$ & $0.2476\pm0.0090$ & $-$ \\ 
    Argument of periastron (deg)\dotfill & $\omega_\mathrm{b}$ & $102.0\pm1.6$ & $-$ \\ 
    Semi-major axis (au)\dotfill & $a_\mathrm{b}$ & $0.0688\pm0.0017$ & $-$ \\ 
    Semi-major axis over host radius\dotfill & $a_\mathrm{b}/R_\star$ & $13.34\pm0.30$ & $-$ \\ 
    Inclination (deg)\dotfill & $i_\mathrm{b}$ & $87.59\pm0.34$ & $-$ \\ 
    Impact parameter \dotfill & $b_\mathrm{tra;b}$ & $0.425\pm0.055$ & $-$ \\ 
    Total transit duration (hours)\dotfill & $T_\mathrm{tot;b}$ & $2.961_{-0.049}^{+0.053}$ & $-$ \\ 
    Companion density (cgs)\dotfill & $\rho_\mathrm{b}$ & $156_{-14}^{+16}$ & $-$ \\ 
    Equilibrium temperature (K)\dotfill & $T_\mathrm{eq;b}$ & $1095\pm17$ & $-$ \\ 
    Transit depth (ppm)\dotfill & $\delta_\mathrm{tr; undil; b; TESS}$ & $6740_{-180}^{+210}$  & $-$ \\ 
    Linear limb-darkening coefficient for TESS\dotfill & $u_\mathrm{1; TESS}$ & $0.19_{-0.13}^{+0.21}$ & $-$ \\ 
    Linear limb-darkening coefficient for TESS\dotfill & $u_\mathrm{2; TESS}$ & $0.30_{-0.30}^{+0.27}$ & $-$ \\ 
    Linear limb-darkening coefficient for NEID\dotfill & $u_\mathrm{1; NEID}$ & $0.85\pm0.55$ & $-$ \\ 
    Linear limb-darkening coefficient for NEID\dotfill & $u_\mathrm{2; NEID}$ & $-0.13_{-0.43}^{+0.50}$ & $-$ \\[0.1cm]
    \hline
    \hline 
    \end{tabular}
    \label{tab:allesfitter_results}
\end{table*}

\section{Dynamical Timescales}\label{sec:dyntimescales}

Interactions with the primary star in a two-body system may provide a damping mechanism to push a misaligned or eccentric orbit toward its lowest energy state \citep[e.g.][]{2008ApJ...678.1396J, 2010ApJ...725.1995M}. In this section, we examine relevant dynamical timescales and discuss their implications for the evolutionary history of TOI-2533 $b$. 

\subsection{Tidal Realignment and Turbulent Friction}\label{sec:tauCE}

Cross-comparing with Figure 9 from \cite{2022ApJ...927...22S}, we conclude that, based on the host star's effective temperature $T_{\rm eff} = 6180\pm60$ K and its metallicity ${\rm [Fe/H]} = -0.3\pm0.2$ as reported in \cite{2023AJ....166..225S}, TOI-2533 most likely hosts a convective envelope and lies below the Kraft break \citep{1967ApJ...150..551K}. We note that TOI-2533's projected stellar rotational velocity $v\sin i_*=8.4^{+1.5}_{-1.4}$ km/s is consistent with the star lying either below or above the Kraft break (see e.g., \citealt{2021AJ....161...68L}). Stars with convective envelopes have efficient magnetic dynamos at a young age -- especially relevant for star-disc interactions over timescales of a few million years -- which may facilitate orbital realignment \citep{2015ApJ...811...82S}. 

Furthermore, star-companion tidal interactions push systems toward low obliquities over time. Stars above the Kraft break possess radiative envelopes and shallower convective zones, typically resulting in longer predicted tidal realignment timescales within the equilibrium tides framework \citep[e.g.][]{2012ApJ...757...18A, 2022ApJ...926L..17R}. For hot Jupiters, realignment may also be induced through a coupling of the companion's orbit and the host star's oscillation modes (i.e., gravity modes in the stellar interior) that preferentially realign systems with cool stellar hosts \citep{2024arXiv240305616Z}; however, this mechanism does not extend to systems with companion masses greater than a few Jupiter masses.

Since TOI-2533 falls within a temperature range directly coinciding with the Kraft break transition, we consider both the case in which it does and does not have a substantive convective envelope. For stars with significant convective envelopes, i.e., those below the Kraft break, the empirical tidal realignment timescale can be estimated within the equilibrium tides framework as \citep{2012ApJ...757...18A}

\begin{equation}
    \tau_{\rm CE} = 10^{10} q^{-2} \left(\frac{a/R_\star}{40}\right)^6,
    \label{eq:tauCE}
\end{equation}
calibrated from observations made in binary star systems by \cite{1977A&A....57..383Z}. Here, $q \equiv M_{\rm b}/M_*$ represents the mass ratio of the companion to the host star, and $a/R_\star$ denotes the companion's orbital separation scaled by the stellar radius. Hotter stars with radiative envelopes follow a separate empirical scaling given as

\begin{equation}
    \tau_{\rm RA} = 6.25\times10^{9} q^{-2} \left(1 + q\right)^{-5/6} \left(\frac{a/R_\star}{6}\right)^{17/2}.
\end{equation}

Based on our derived properties for the TOI-2533 system, we find that $\tau_{\rm CE} \sim 10^9$ years and $\tau_{\rm RA} \sim10^{15}$ years. If the host star possesses a substantial convective envelope, the system's age and $\tau_{\rm CE}$ coincide within a 3$\sigma$ margin ($\Gamma = 4.26^{+2.18}_{-1.74}$ Gyr; \citealt{2023AJ....166..225S}), such that the companion may have plausibly realigned over its lifetime if it formed in or was previously excited to a misaligned configuration. Alternatively, if the host star has a radiative envelope, the system would be difficult to realign within the system lifetime, implying that the observed alignment is likely primordial.

\subsection{Eccentricity Evolution Driven by Tidal Dissipation}\label{sec:eccevolution}

For comparison with the system's tidal realignment timescale, we also consider the evolution timescale of TOI-2533 $b$'s orbital eccentricity $\tau_e \sim e/\dot{e}$. Assuming that the brown dwarf is in pseudo-synchronous rotation -- i.e., that its rotational frequency $\Omega$ is commensurate with its orbital angular frequency at periastron (although see \citet{2024arXiv240306979D}, which describes possible expected deviations from pseudo-synchronisation) -- its energy dissipation rate can be expressed as

\begin{equation}
    \dot{e} = \left[\frac{21 k_2 GM_*^2 \Omega R_*^5 \zeta(e)}{2\mathcal{Q}a^6}\right] \frac{a(1-e^2)}{GM_*M_{b}e},
\end{equation}

where $M_*$ and $R_*$ correspond to the mass and radius of the star, and $M_b$ refers to the companion's mass. Both $\Omega$ and the corrective factor $\zeta(e)$ depend on the companion's eccentricity, with each given as \citep{1981A&A....99..126H, 2008Icar..193..637W}

\begin{equation}
    \Omega = \frac{1}{P}\left[\frac{f_1(e)}{f_2(e) \beta^3}\right],
\end{equation}

\begin{equation}
    \zeta(e) = \frac{2}{7}\left[\frac{f_0(e)}{\beta^{15}} - 2 \frac{f_1(e)}{\beta^{12}} + \frac{f_2(e)}{\beta^9}\right]
\end{equation}

where

\begin{eqnarray}
    f_0(e) &=& 1 + \frac{31}{2}e^2 + \frac{255}{8}e^4 + \frac{185}{16}e^6 + \frac{25}{64}e^8,\\
    f_1(e) &=& 1 + \frac{15}{2}e^2 + \frac{45}{8}e^4 + \frac{5}{16}e^6, \\
    f_2(e) &=& 1 + 3e^2 + \frac{3}{8}e^4, \mbox{ and}, \\
    \beta &\equiv& \sqrt{1-e^2}.
\end{eqnarray}

Following \cite{2022ApJ...926L..17R}, we set fiducial values $\mathcal{Q} = 10^5$ as the companion's effective tidal dissipation parameter and $k_2 = 0.3$ as the Love number, which represents the companion's rigidity and susceptibility of its shape to change in response to a tidal potential.

For TOI-2533 $b$, we estimate $\tau_e \sim 10^{12}$ years, which exceeds the proper system's age by three orders of magnitude. Therefore, we anticipate that, if TOI-2533 has a substantive convective envelope, the system should tidally realign before its orbital circularisation.

\section{Discussion}\label{sec:discussion}

\subsection{Comparison of TOI-2533 $b$ with Theoretical Models for Brown Dwarfs}\label{sec:massradius}


Fully convective objects with Kelvin-Helmholtz timescales on the order of a few million years were postulated around sixty years ago by \cite{1963ApJ...137.1121K, 1963ApJ...137.1126K}, and despite the discovery of over 4000 brown dwarfs so far\footnote{As of February 2024; see the {UltracoolSheet} v2.0.0 catalogue \citep{UltracoolSheet_v2}.}, ongoing debate continues to refine our understanding of their formation mechanisms within the context of star and planet formation. Such objects fall short of initiating hydrogen nuclear fusion, with upper masses spanning $M \sim M_{\rm HBMM} = 0.07-0.09$ M$_{\odot}$ \citep{2001RvMP...73..719B, 2019ApJ...871..227F, 2023A&A...671A.119C}, but are massive enough for deuterium burning in their cores ($M > M_{\rm DBMM} = 11-16.3$ M$_{\rm \tiny Jup}$; \citealt{2011ApJ...727...57S}). Relatively high-mass brown dwarfs ($\gtrapprox60$ M$_{\rm \tiny Jup}$) have radii predominantly regulated by electron-degeneracy pressure, while those at the lower limit ($\sim 14-20$ M$_{\rm \tiny Jup}$) are mainly constrained by Coulomb pressure \citep{2006AREPS..34..193B}.

\begin{figure}[h]
    \centering
    \includegraphics[width = \linewidth]{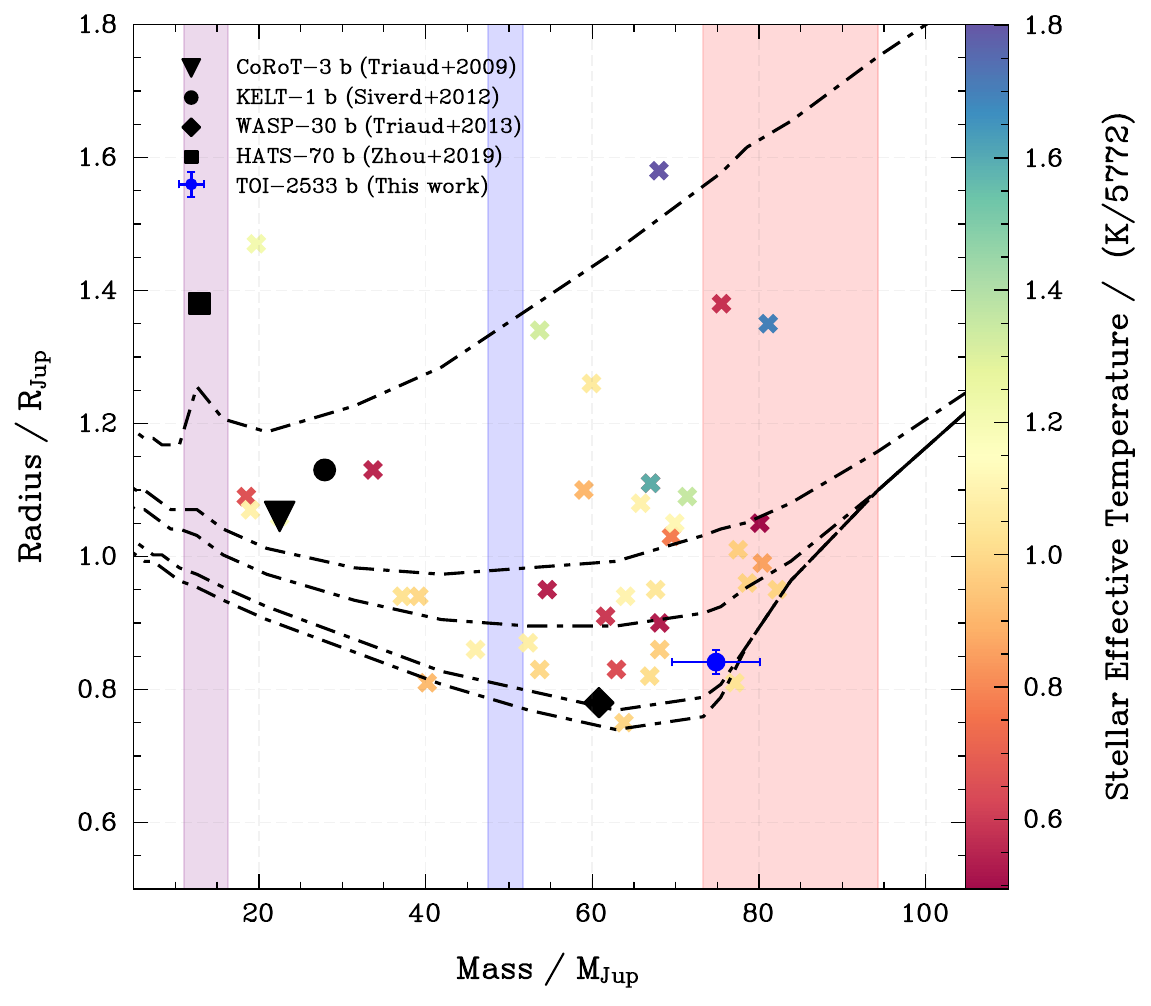}
    \caption{Mass-radius diagram for the sample of known transiting brown dwarfs as of September 2023 \citep{2023MNRAS.519.5177C}. The known transiting brown dwarfs to date with measured spin-orbit angles are indicated as black symbols. COND isochrones with ages 0.1, 0.5, 1, 5, and 10 Gyr, neglecting dust opacity in the radiative transfer equation, are indicated as dashed lines (top to bottom; \citealt{2003A&A...402..701B}). For reference, the age of the TOI-2533 system (indicated as a blue dot) is estimated as $4.26^{+2.18}_{-1.74}$ Gyr \citep{2023AJ....166..225S}. Colours indicate the stellar effective temperature scaled by the Sun's ($T_{\rm eff} = 5772$ K), and the shaded regions (left to right) show the boundaries for the deuterium-burning minimum mass (purple), lithium-burning minimum mass (blue), and hydrogen-burning minimum mass (pink).}
    \label{fig:mass_radius}
\end{figure}

The mass of TOI-2533 $b$ ($M_b = 74.9\pm5.3$ M$_{\rm \tiny Jup} = 0.0715\pm0.0051$ M$_\odot$) falls marginally within the hydrogen-burning minimum mass regime ($0.07-0.09$ M$_\odot$), leaving its classification ambiguous between a brown dwarf and an ultra-cool dwarf star (see Figure \ref{fig:mass_radius}). Notably, TOI-2533 $b$'s mass falls above the lithium-burning minimum mass limit ($M_{\rm LBMM} \sim 51.4$ M$_{\rm \tiny Jup}$ \citealt{2022MNRAS.510.2841M}). Nevertheless, lithium absorption features are not expected in the brown dwarf's spectrum, since lithium is expected to be completely depleted in brown dwarf atmospheres older than $100-250$ Myr (e.g., \citealt{2000ARA&A..38..485B, 2008ApJ...689.1295K}). 

Evolutionary models for field brown dwarfs suggest that objects with ages around $0.5$ Gyr exhibit radii comparable to those of Jovian planets, while high-mass brown dwarfs ($\geq$ 50 M$_{\rm \tiny Jup}$) tend to shrink over time, resulting in radii smaller than $1$ R$_{\rm \tiny Jup}$ after $1$ Gyr \citep{2003A&A...402..701B, 2011ApJ...736...47B, 2023A&A...671A.119C}. This contrasts with the straightforward mass-based pattern observed in stars ($R \propto M^\xi$, with $\xi\in\mathbb{R}^+$; \citealt{2013sse..book.....K}). The estimated radius of TOI-2533 $b$ is $0.841\pm0.018$ R$_{\rm \tiny Jup}$ based on our TESS light curve modelling (see Section \ref{sec:spin-orbit_modelling}), consistent with theoretical predictions given the object's mass ($M_b = 74.9\pm5.3$ M$_{\rm \tiny Jup}$) and age ($4.26^{+2.18}_{-1.74}$ Gyr; see Figure \ref{fig:mass_radius}).

\subsection{Possible Formation Scenarios for TOI-2533 $b$}\label{sec:formationscenarios}

At first glance, TOI-2533 $b$ appears similar to a scaled-up hot Jupiter system, with a very short orbital period ($P \sim 6.68$ days) but higher mass ($M \sim 75~M_{\rm \tiny Jup}$). Short-period brown dwarfs around Sun-like stars also have a very low projected occurrence rate ($\lesssim2\%$) comparable to the $\sim1\%$ occurrence rate of hot Jupiters \citep{2019A&A...631A.125K, 2023ASPC..534..275O}. While the brown dwarf's aligned and moderately eccentric orbit may be suggestive of mechanisms such as co-planar high-eccentricity migration \citep{2015ApJ...805...75P}, no outer planetary companion is known to date, and the brown dwarf's eccentricity evolution is slow compared to the system's age (see Section \ref{sec:dyntimescales}). Therefore, formation through high-eccentricity migration, as has been suggested for hot Jupiters based on their stellar obliquity distribution \citep{2022ApJ...926L..17R}, is unlikely for this system. 

The high mass of TOI-2533 $b$, together with the system's relatively high mass ratio (see Table \ref{tab:allesfitter_results}), instead suggests a formation process reminiscent of stellar binary systems \citep{2013ARA&A..51..269D}. Several mechanisms have been proposed to explain the formation of close binary systems, including gravitational capture \citep[e.g.][]{1975MNRAS.172P..15F, 2012Ap&SS.341..395K}, turbulent fragmentation \citep[e.g.][]{2010ApJ...725.1485O, 2011ASPC..447...47K}, and disc fragmentation \citep[e.g.][]{1953ApJ...118..513H, 1994MNRAS.271..999B}. 

Gravitational capture, in which a companion is captured into a bound orbit by the gravitational well of the primary star, typically occurs in dense environments \citep[e.g.][and references therein]{1975MNRAS.172P..15F, 1991MNRAS.249..584C, 2008gady.book.....B, 2012Ap&SS.341..395K}. In globular clusters, the capture rate of brown dwarfs by main-sequence stars has been estimated with an upper limit of 15\% based on the absence of observed transits in the 47 Tucan{\ae} globular cluster \citep{gilliland2000lack, 2003MNRAS.343L..53B}, whereas the capture rate should be lower for brown dwarfs born in loose stellar clusters with a comparatively low number density  \citep{2003astro.ph..7280G}. Thus, while somewhat rare, gravitational capture may be consistent with the low occurrence rate of short-period brown dwarfs around Sun-like stars \citep[$\lesssim2\%$;][]{2019A&A...631A.125K, 2023ASPC..534..275O} and remains as a possibility for the formation of TOI-2533 $b$.

In the framework of turbulent fragmentation, density perturbations within a molecular core already under collapse induce fragmentation into smaller cores or filaments when their Jeans mass is exceeded, leading to the formation of gravitationally bound binary or multiple body systems \citep{2002ApJ...576..870P, 2004A&A...414..633G, 2010ApJ...725.1485O, 2019A&A...628A.112K, 2023ASPC..534..275O}. Turbulent fragmentation is typically expected to produce no preferential directionality of spin-orbit orientations due to the stochasticity of angular momenta throughout the molecular cloud. As a result, turbulent fragmentation would naturally produce an isotropic distribution of primordial spin-orbit orientations \citep{2010ApJ...725.1485O, 2017NatAs...1E.172L}. Since TOI-2533 most likely has a convective envelope, and, in that case, the system has a relatively short tidal realignment timescale $\tau_{\rm CE}$, it is possible that the system may have formed through turbulent fragmentation and subsequently realigned to its currently observed orbital configuration.

Disc fragmentation during the second collapse phase of a molecular cloud\footnote{In the first collapse phase, also known as the isothermal phase, the number density around the centre of the cloud is on the order of $n_c \leq 10^{11}$ cm$^{-3}$, while the second collapse occurs when $n_c>10^{16}$ cm$^{-3}$ and the cloud reaches a temperature of about $2000$ K, causing molecular H$_{2}$ dissociation and a reduction of the adiabatic constant $\gamma \equiv C_{P}/C_{V}$ to $\gamma_{\rm critical} < 4/3$ \citep{1969MNRAS.145..271L}.} is a third potential pathway to form close stellar binaries \citep{1999ApJ...523L.155T, 2006A&A...458..817W, 2012ApJ...760...99P, 2020MNRAS.491.5158T}. In this scenario, a rotating molecular cloud flattens into a disc-like structure before its fragmentation (see Figure 1 in \citealt{1980ApJ...242..209B}, and \citealt{2009Sci...323..754K, 2011MNRAS.417.2036B}), which takes place when there is a sufficiently high ratio between rotational and gravitational energies during the free-fall collapse phase \citep{2018MNRAS.480.4434W}. These fragments break into smaller ones, increasing density, opacity, and temperature until further fragmentation ceases \citep{2009itss.book.....P}. Accretion of material onto these proto-brown dwarfs continues until their surrounding gas dissipates and before they reach stellar masses \citep{2002MNRAS.332L..65B, Umbreit2005The}. Disc fragmentation naturally produces aligned and moderately eccentric close-in systems like TOI-2533 $b$ \citep{1994AJ....107..306H, 1994ApJ...436..335L, 1994MNRAS.271..999B, 2010ApJ...708.1585K}, with no need to invoke tidal realignment. 

While the formation mechanism for TOI-2533 b remains ambiguous across these three mechanisms, the system highlights the need for further brown dwarf spin-orbit constraints. In particular, brown dwarfs with long projected tidal realignment timescales may inform how the broader population of short-period brown dwarfs, including TOI-2533 b, likely reached their tight orbits. An extended sample of brown dwarf spin-orbit orientations has the potential to more generally inform the dynamical pathways that sculpt main-sequence/brown dwarf binaries.

\section{Conclusions}\label{sec:conclusions}

TOI-2533 $b$ stands out as the fifth transiting brown dwarf with a measured spin-orbit angle reported to date, as well as the twelfth result of the SOLES survey. We have leveraged in-transit radial velocity observations from the NEID/WIYN spectrograph to show that TOI-2533 $b$'s orbit is consistent with alignment, with a low-sky-projected spin-orbit angle, $\lambda = -7\pm14^\circ$, that falls within the range observed for quiescent near-resonant exoplanet systems \citep{2023AJ....166..266R} -- a possible tracer of primordial circumstellar disc alignments. 

With its aligned and moderately eccentric orbit, TOI-2533 $b$'s origin is consistent with stellar binary-like formation through disc fragmentation. This formation pathway is consistent with the brown dwarf's relatively high mass $M_b = 74.9\pm5.3$ M$_{\rm Jup}$ and relatively high mass ratio $q = 0.0702_{-0.0019}^{+0.0021}$. However, we cannot rule out the possibility that the system initially had a larger misalignment that was corrected over its lifetime, given the short tidal realignment timescale $\tau_{\rm CE}$ in the case that the host star has a significant convective envelope. Further brown dwarf spin-orbit measurements across a wider parameter space, particularly for systems with long projected tidal realignment timescales, would offer useful insights to delineate the range of orbital architectures for these intermediate-mass eclipsing systems.

\vspace{5mm}

\section*{Acknowledgements}

T.F. acknowledges support from Yale Graduate School of Arts and Sciences. M.R. acknowledges support from Heising-Simons Foundation Grant \#2023-4478, Oracle for Research grant No. CPQ-3033929, and JPL/NASA Grant \#1709150. S.W. acknowledges support from the Heising-Simons Foundation Grant $\#$2023-4050. We acknowledge support from the NASA Exoplanets Research Program NNH23ZDA001N-XRP (Grant No. 80NSSC24K0153). 
This paper contains data taken with the NEID instrument, which was funded by the NASA-NSF Exoplanet Observational Research (NN-EXPLORE) partnership and built by Pennsylvania State University. NEID is installed on the WIYN telescope, which is operated by the National Optical Astronomy Observatory, and the NEID archive is operated by the NASA Exoplanet Science Institute at the California Institute of Technology. NN-EXPLORE is managed by the Jet Propulsion Laboratory, California Institute of Technology under contract with the National Aeronautics and Space Administration. All {\it TESS} data used in this paper can be downloaded from MAST: \dataset[10.17909/vpkg-8e10]{http://dx.doi.org/10.17909/vpkg-8e10}. 

\facilities{NEID, TESS, TRES, The Encyclop{\ae}dia of Exoplanetary Systems, TEPCat.}

\software{\texttt{numpy} \citep{2020Natur.585..357H}, \texttt{pandas} \citep{2022zndo...3509134T}, \texttt{astropy} \citep{astropy:2013, astropy:2018, astropy:2022}, \texttt{matplotlib} \citep{2007CSE.....9...90H}, \texttt{smplotlib} \citep{jiaxuan_li_2023_8126529}, \texttt{lightkurve} \citep{2018ascl.soft12013L}, , \texttt{allesfitter} \citep{2021ApJS..254...13G}, \texttt{dynesty} \citep{2020MNRAS.493.3132S}, \texttt{tpfplotter} \citep{2020A&A...635A.128A}, \texttt{VOSA} \citep{2008A&A...492..277B}.}

\appendix

\section{No Bound Stellar Companions to TOI-2533}\label{app:noboundcompanion}

An inspection of the TESS Target Pixel Files (TPF) using the {\sc tpfplotter}\footnote{\url{https://github.com/jlillo/tpfplotter}} \citep{2020A&A...635A.128A} software initially revealed two nearby sources with $\Delta G \leq 6$ mag relative to TOI-2533: {\it (i)} \emph{Gaia} DR3 1259922127931777664 ($\mu_\alpha = -1.62\pm0.02$ mas/yr, $\mu_\delta = 2.91\pm0.02$ mas/yr, $\overline\pi = 0.025\pm0.023$ mas) and {\it (ii)} \emph{Gaia} DR3 1259921754269906560 ($\mu_\alpha = -12.50\pm0.02$ mas/yr, $\mu_\delta = 15.62\pm0.01$ mas/yr, $\overline\pi = 1.71\pm0.02$ mas).

The 3D velocity difference between \emph{Gaia} DR3 1259921754269906560 and TOI-2533 is $\Delta v_{\rm 3D} \sim 13.77$ km/s based on the two stars' parallax and proper motion measurements, surpassing the threshold of $\Delta v_{\rm 3D} \leq 2$ km/s typically associated with gravitationally bound objects with $a < 10^7$ au \citep{2019ApJ...884..173K}. \emph{Gaia} DR3 1259921754269906560 maintains a projected separation of $53.07\arcsec$ from TOI-2533, corresponding to approximately $19600$ AU. While the projected angular separation between TOI-2533 and \emph{Gaia} DR3 1259922127931777664 is smaller, at $42.05\arcsec$, the projected distance between the two stars is much larger, at $d \sim 38910\pm35579$ pc from inverting the parallax of \emph{Gaia} DR3 (although we caution that this approach overestimates large distances; see e.g., \citealt{2021A&A...647A.169H}). By employing corrections from \cite{2021AJ....161..147B}, the median geometric distance is found to be $d \sim 11422_{-1762}^{+2613}$ pc. Therefore, the likelihood that either star is gravitationally bound to TOI-2533 is low. 

Neither of these two sources nor TOI-2533 itself is present in the catalogue of binary systems presented by \cite{2021MNRAS.506.2269E}. We also note that no close-separation companions have been identified within a $\sim8\arcsec$ field of view from high-resolution adaptive optics imaging on the Exoplanet Follow-up Observing Program (ExoFOP) website\footnote{\url{https://exofop.ipac.caltech.edu/tess/}}.

\bibliography{bib}{}

\begin{thebibliography}{}
\expandafter\ifx\csname natexlab\endcsname\relax\def\natexlab#1{#1}\fi
\providecommand{\url}[1]{\href{#1}{#1}}
\providecommand{\dodoi}[1]{doi:~\href{http://doi.org/#1}{\nolinkurl{#1}}}
\providecommand{\doeprint}[1]{\href{http://ascl.net/#1}{\nolinkurl{http://ascl.net/#1}}}
\providecommand{\doarXiv}[1]{\href{https://arxiv.org/abs/#1}{\nolinkurl{https://arxiv.org/abs/#1}}}

\bibitem[{{Albrecht} {et~al.}(2013){Albrecht}, {Setiawan}, {Torres}, {Fabrycky}, \& {Winn}}]{2013ApJ...767...32A}
{Albrecht}, S., {Setiawan}, J., {Torres}, G., {Fabrycky}, D.~C., \& {Winn}, J.~N. 2013, \apj, 767, 32, \dodoi{10.1088/0004-637X/767/1/32}

\bibitem[{{Albrecht} {et~al.}(2012){Albrecht}, {Winn}, {Johnson}, {Howard}, {Marcy}, {Butler}, {Arriagada}, {Crane}, {Shectman}, {Thompson}, {Hirano}, {Bakos}, \& {Hartman}}]{2012ApJ...757...18A}
{Albrecht}, S., {Winn}, J.~N., {Johnson}, J.~A., {et~al.} 2012, \apj, 757, 18, \dodoi{10.1088/0004-637X/757/1/18}

\bibitem[{{Albrecht} {et~al.}(2022){Albrecht}, {Dawson}, \& {Winn}}]{2022PASP..134h2001A}
{Albrecht}, S.~H., {Dawson}, R.~I., \& {Winn}, J.~N. 2022, \pasp, 134, 082001, \dodoi{10.1088/1538-3873/ac6c09}

\bibitem[{{Aller} {et~al.}(2020){Aller}, {Lillo-Box}, {Jones}, {Miranda}, \& {Barcel{\'o} Forteza}}]{2020A&A...635A.128A}
{Aller}, A., {Lillo-Box}, J., {Jones}, D., {Miranda}, L.~F., \& {Barcel{\'o} Forteza}, S. 2020, \aap, 635, A128, \dodoi{10.1051/0004-6361/201937118}

\bibitem[{{Astropy Collaboration} {et~al.}(2013){Astropy Collaboration}, {Robitaille}, {Tollerud}, {Greenfield}, {Droettboom}, {Bray}, {Aldcroft}, {Davis}, {Ginsburg}, {Price-Whelan}, {Kerzendorf}, {Conley}, {Crighton}, {Barbary}, {Muna}, {Ferguson}, {Grollier}, {Parikh}, {Nair}, {Unther}, {Deil}, {Woillez}, {Conseil}, {Kramer}, {Turner}, {Singer}, {Fox}, {Weaver}, {Zabalza}, {Edwards}, {Azalee Bostroem}, {Burke}, {Casey}, {Crawford}, {Dencheva}, {Ely}, {Jenness}, {Labrie}, {Lim}, {Pierfederici}, {Pontzen}, {Ptak}, {Refsdal}, {Servillat}, \& {Streicher}}]{astropy:2013}
{Astropy Collaboration}, {Robitaille}, T.~P., {Tollerud}, E.~J., {et~al.} 2013, \aap, 558, A33, \dodoi{10.1051/0004-6361/201322068}

\bibitem[{{Astropy Collaboration} {et~al.}(2018){Astropy Collaboration}, {Price-Whelan}, {Sip{\H{o}}cz}, {G{\"u}nther}, {Lim}, {Crawford}, {Conseil}, {Shupe}, {Craig}, {Dencheva}, {Ginsburg}, {Vand erPlas}, {Bradley}, {P{\'e}rez-Su{\'a}rez}, {de Val-Borro}, {Aldcroft}, {Cruz}, {Robitaille}, {Tollerud}, {Ardelean}, {Babej}, {Bach}, {Bachetti}, {Bakanov}, {Bamford}, {Barentsen}, {Barmby}, {Baumbach}, {Berry}, {Biscani}, {Boquien}, {Bostroem}, {Bouma}, {Brammer}, {Bray}, {Breytenbach}, {Buddelmeijer}, {Burke}, {Calderone}, {Cano Rodr{\'\i}guez}, {Cara}, {Cardoso}, {Cheedella}, {Copin}, {Corrales}, {Crichton}, {D'Avella}, {Deil}, {Depagne}, {Dietrich}, {Donath}, {Droettboom}, {Earl}, {Erben}, {Fabbro}, {Ferreira}, {Finethy}, {Fox}, {Garrison}, {Gibbons}, {Goldstein}, {Gommers}, {Greco}, {Greenfield}, {Groener}, {Grollier}, {Hagen}, {Hirst}, {Homeier}, {Horton}, {Hosseinzadeh}, {Hu}, {Hunkeler}, {Ivezi{\'c}}, {Jain}, {Jenness}, {Kanarek}, {Kendrew}, {Kern}, {Kerzendorf}, {Khvalko}, {King}, {Kirkby}, {Kulkarni},
  {Kumar}, {Lee}, {Lenz}, {Littlefair}, {Ma}, {Macleod}, {Mastropietro}, {McCully}, {Montagnac}, {Morris}, {Mueller}, {Mumford}, {Muna}, {Murphy}, {Nelson}, {Nguyen}, {Ninan}, {N{\"o}the}, {Ogaz}, {Oh}, {Parejko}, {Parley}, {Pascual}, {Patil}, {Patil}, {Plunkett}, {Prochaska}, {Rastogi}, {Reddy Janga}, {Sabater}, {Sakurikar}, {Seifert}, {Sherbert}, {Sherwood-Taylor}, {Shih}, {Sick}, {Silbiger}, {Singanamalla}, {Singer}, {Sladen}, {Sooley}, {Sornarajah}, {Streicher}, {Teuben}, {Thomas}, {Tremblay}, {Turner}, {Terr{\'o}n}, {van Kerkwijk}, {de la Vega}, {Watkins}, {Weaver}, {Whitmore}, {Woillez}, {Zabalza}, \& {Astropy Contributors}}]{astropy:2018}
{Astropy Collaboration}, {Price-Whelan}, A.~M., {Sip{\H{o}}cz}, B.~M., {et~al.} 2018, \aj, 156, 123, \dodoi{10.3847/1538-3881/aabc4f}

\bibitem[{{Astropy Collaboration} {et~al.}(2022){Astropy Collaboration}, {Price-Whelan}, {Lim}, {Earl}, {Starkman}, {Bradley}, {Shupe}, {Patil}, {Corrales}, {Brasseur}, {N{"o}the}, {Donath}, {Tollerud}, {Morris}, {Ginsburg}, {Vaher}, {Weaver}, {Tocknell}, {Jamieson}, {van Kerkwijk}, {Robitaille}, {Merry}, {Bachetti}, {G{"u}nther}, {Aldcroft}, {Alvarado-Montes}, {Archibald}, {B{'o}di}, {Bapat}, {Barentsen}, {Baz{'a}n}, {Biswas}, {Boquien}, {Burke}, {Cara}, {Cara}, {Conroy}, {Conseil}, {Craig}, {Cross}, {Cruz}, {D'Eugenio}, {Dencheva}, {Devillepoix}, {Dietrich}, {Eigenbrot}, {Erben}, {Ferreira}, {Foreman-Mackey}, {Fox}, {Freij}, {Garg}, {Geda}, {Glattly}, {Gondhalekar}, {Gordon}, {Grant}, {Greenfield}, {Groener}, {Guest}, {Gurovich}, {Handberg}, {Hart}, {Hatfield-Dodds}, {Homeier}, {Hosseinzadeh}, {Jenness}, {Jones}, {Joseph}, {Kalmbach}, {Karamehmetoglu}, {Ka{l}uszy{'n}ski}, {Kelley}, {Kern}, {Kerzendorf}, {Koch}, {Kulumani}, {Lee}, {Ly}, {Ma}, {MacBride}, {Maljaars}, {Muna}, {Murphy}, {Norman}, {O'Steen},
  {Oman}, {Pacifici}, {Pascual}, {Pascual-Granado}, {Patil}, {Perren}, {Pickering}, {Rastogi}, {Roulston}, {Ryan}, {Rykoff}, {Sabater}, {Sakurikar}, {Salgado}, {Sanghi}, {Saunders}, {Savchenko}, {Schwardt}, {Seifert-Eckert}, {Shih}, {Jain}, {Shukla}, {Sick}, {Simpson}, {Singanamalla}, {Singer}, {Singhal}, {Sinha}, {Sip{H{o}}cz}, {Spitler}, {Stansby}, {Streicher}, {{{S}}umak}, {Swinbank}, {Taranu}, {Tewary}, {Tremblay}, {Val-Borro}, {Van Kooten}, {Vasovi{'c}}, {Verma}, {de Miranda Cardoso}, {Williams}, {Wilson}, {Winkel}, {Wood-Vasey}, {Xue}, {Yoachim}, {Zhang}, {Zonca}, \& {Astropy Project Contributors}}]{astropy:2022}
{Astropy Collaboration}, {Price-Whelan}, A.~M., {Lim}, P.~L., {et~al.} 2022, \apj, 935, 167, \dodoi{10.3847/1538-4357/ac7c74}

\bibitem[{{Bailer-Jones} {et~al.}(2021){Bailer-Jones}, {Rybizki}, {Fouesneau}, {Demleitner}, \& {Andrae}}]{2021AJ....161..147B}
{Bailer-Jones}, C.~A.~L., {Rybizki}, J., {Fouesneau}, M., {Demleitner}, M., \& {Andrae}, R. 2021, \aj, 161, 147, \dodoi{10.3847/1538-3881/abd806}

\bibitem[{{Baraffe} {et~al.}(2003){Baraffe}, {Chabrier}, {Barman}, {Allard}, \& {Hauschildt}}]{2003A&A...402..701B}
{Baraffe}, I., {Chabrier}, G., {Barman}, T.~S., {Allard}, F., \& {Hauschildt}, P.~H. 2003, \aap, 402, 701, \dodoi{10.1051/0004-6361:20030252}

\bibitem[{{Basri}(2000)}]{2000ARA&A..38..485B}
{Basri}, G. 2000, \araa, 38, 485, \dodoi{10.1146/annurev.astro.38.1.485}

\bibitem[{{Basri} \& {Brown}(2006)}]{2006AREPS..34..193B}
{Basri}, G., \& {Brown}, M.~E. 2006, Annual Review of Earth and Planetary Sciences, 34, 193, \dodoi{10.1146/annurev.earth.34.031405.125058}

\bibitem[{{Bate}(2011)}]{2011MNRAS.417.2036B}
{Bate}, M.~R. 2011, \mnras, 417, 2036, \dodoi{10.1111/j.1365-2966.2011.19386.x}

\bibitem[{{Bate} {et~al.}(2002){Bate}, {Bonnell}, \& {Bromm}}]{2002MNRAS.332L..65B}
{Bate}, M.~R., {Bonnell}, I.~A., \& {Bromm}, V. 2002, \mnras, 332, L65, \dodoi{10.1046/j.1365-8711.2002.05539.x}

\bibitem[{{Bate} {et~al.}(2003){Bate}, {Bonnell}, \& {Bromm}}]{2003MNRAS.339..577B}
---. 2003, \mnras, 339, 577, \dodoi{10.1046/j.1365-8711.2003.06210.x}

\bibitem[{{Bate} {et~al.}(2010){Bate}, {Lodato}, \& {Pringle}}]{2010MNRAS.401.1505B}
{Bate}, M.~R., {Lodato}, G., \& {Pringle}, J.~E. 2010, \mnras, 401, 1505, \dodoi{10.1111/j.1365-2966.2009.15773.x}

\bibitem[{{Bayo} {et~al.}(2008){Bayo}, {Rodrigo}, {Barrado Y Navascu{\'e}s}, {Solano}, {Guti{\'e}rrez}, {Morales-Calder{\'o}n}, \& {Allard}}]{2008A&A...492..277B}
{Bayo}, A., {Rodrigo}, C., {Barrado Y Navascu{\'e}s}, D., {et~al.} 2008, \aap, 492, 277, \dodoi{10.1051/0004-6361:200810395}

\bibitem[{{Benitez} {et~al.}(2014){Benitez}, {Dupke}, {Moles}, {Sodre}, {Cenarro}, {Marin-Franch}, {Taylor}, {Cristobal}, {Fernandez-Soto}, {Mendes de Oliveira}, {Cepa-Nogue}, {Abramo}, {Alcaniz}, {Overzier}, {Hernandez-Monteagudo}, {Alfaro}, {Kanaan}, {Carvano}, {Reis}, {Martinez Gonzalez}, {Ascaso}, {Ballesteros}, {Xavier}, {Varela}, {Ederoclite}, {Vazquez Ramio}, {Broadhurst}, {Cypriano}, {Angulo}, {Diego}, {Zandivarez}, {Diaz}, {Melchior}, {Umetsu}, {Spinelli}, {Zitrin}, {Coe}, {Yepes}, {Vielva}, {Sahni}, {Marcos-Caballero}, {Kitaura}, {Maroto}, {Masip}, {Tsujikawa}, {Carneiro}, {Gonzalez Nuevo}, {Carvalho}, {Reboucas}, {Carvalho}, {Abdalla}, {Bernui}, {Pigozzo}, {Ferreira}, {Chandrachani Devi}, {Bengaly}, {Campista}, {Amorim}, {Asari}, {Bongiovanni}, {Bonoli}, {Bruzual}, {Cardiel}, {Cava}, {Cid Fernandes}, {Coelho}, {Cortesi}, {Delgado}, {Diaz Garcia}, {Espinosa}, {Galliano}, {Gonzalez-Serrano}, {Falcon-Barroso}, {Fritz}, {Fernandes}, {Gorgas}, {Hoyos}, {Jimenez-Teja}, {Lopez-Aguerri}, {Lopez-San Juan},
  {Mateus}, {Molino}, {Novais}, {OMill}, {Oteo}, {Perez-Gonzalez}, {Poggianti}, {Proctor}, {Ricciardelli}, {Sanchez-Blazquez}, {Storchi-Bergmann}, {Telles}, {Schoennell}, {Trujillo}, {Vazdekis}, {Viironen}, {Daflon}, {Aparicio-Villegas}, {Rocha}, {Ribeiro}, {Borges}, {Martins}, {Marcolino}, {Martinez-Delgado}, {Perez-Torres}, {Siffert}, {Calvao}, {Sako}, {Kessler}, {Alvarez-Candal}, {De Pra}, {Roig}, {Lazzaro}, {Gorosabel}, {Lopes de Oliveira}, {Lima-Neto}, {Irwin}, {Liu}, {Alvarez}, {Balmes}, {Chueca}, {Costa-Duarte}, {da Costa}, {Dantas}, {Diaz}, {Fabregat}, {Ferrari}, {Gavela}, {Gracia}, {Gruel}, {Gutierrez}, {Guzman}, {Hernandez-Fernandez}, {Herranz}, {Hurtado-Gil}, {Jablonsky}, {Laporte}, {Le Tiran}, {Licandro}, {Lima}, {Martin}, {Martinez}, {Montero}, {Penteado}, {Pereira}, {Peris}, {Quilis}, {Sanchez-Portal}, {Soja}, {Solano}, {Torra}, \& {Valdivielso}}]{2014arXiv1403.5237B}
{Benitez}, N., {Dupke}, R., {Moles}, M., {et~al.} 2014, arXiv e-prints, arXiv:1403.5237, \dodoi{10.48550/arXiv.1403.5237}

\bibitem[{Best {et~al.}(2024)Best, Dupuy, Liu, Sanghi, Siverd, \& Zhang}]{UltracoolSheet_v2}
Best, W. M.~J., Dupuy, T.~J., Liu, M.~C., {et~al.} 2024, The UltracoolSheet: Photometry, Astrometry, Spectroscopy, and Multiplicity for 4000+ Ultracool Dwarfs and Imaged Exoplanets,  Zenodo, \dodoi{10.5281/ZENODO.4169084}

\bibitem[{{Binney} \& {Tremaine}(2008)}]{2008gady.book.....B}
{Binney}, J., \& {Tremaine}, S. 2008, {Galactic Dynamics: Second Edition}

\bibitem[{{Blanco-Cuaresma}(2019)}]{Blanco2019}
{Blanco-Cuaresma}, S. 2019, \mnras, 486, 2075, \dodoi{10.1093/mnras/stz549}

\bibitem[{{Blanco-Cuaresma} {et~al.}(2014){Blanco-Cuaresma}, {Soubiran}, {Heiter}, \& {Jofr{\'e}}}]{Blanco2014}
{Blanco-Cuaresma}, S., {Soubiran}, C., {Heiter}, U., \& {Jofr{\'e}}, P. 2014, \aap, 569, A111, \dodoi{10.1051/0004-6361/201423945}

\bibitem[{{Bodenheimer} {et~al.}(1980){Bodenheimer}, {Tohline}, \& {Black}}]{1980ApJ...242..209B}
{Bodenheimer}, P., {Tohline}, J.~E., \& {Black}, D.~C. 1980, \apj, 242, 209, \dodoi{10.1086/158457}

\bibitem[{{Bonnell} \& {Bate}(1994)}]{1994MNRAS.271..999B}
{Bonnell}, I.~A., \& {Bate}, M.~R. 1994, \mnras, 271, 999, \dodoi{10.1093/mnras/271.4.999}

\bibitem[{{Bonnell} {et~al.}(2008){Bonnell}, {Clark}, \& {Bate}}]{2008MNRAS.389.1556B}
{Bonnell}, I.~A., {Clark}, P., \& {Bate}, M.~R. 2008, \mnras, 389, 1556, \dodoi{10.1111/j.1365-2966.2008.13679.x}

\bibitem[{{Bonnell} {et~al.}(2003){Bonnell}, {Clarke}, {Bate}, {McCaughrean}, {Pringle}, \& {Zinnecker}}]{2003MNRAS.343L..53B}
{Bonnell}, I.~A., {Clarke}, C.~J., {Bate}, M.~R., {et~al.} 2003, \mnras, 343, L53, \dodoi{10.1046/j.1365-8711.2003.06855.x}

\bibitem[{{Boss}(1997)}]{1997Sci...276.1836B}
{Boss}, A.~P. 1997, Science, 276, 1836, \dodoi{10.1126/science.276.5320.1836}

\bibitem[{{Bowler} {et~al.}(2020){Bowler}, {Blunt}, \& {Nielsen}}]{2020AJ....159...63B}
{Bowler}, B.~P., {Blunt}, S.~C., \& {Nielsen}, E.~L. 2020, \aj, 159, 63, \dodoi{10.3847/1538-3881/ab5b11}

\bibitem[{{Burrows} {et~al.}(2011){Burrows}, {Heng}, \& {Nampaisarn}}]{2011ApJ...736...47B}
{Burrows}, A., {Heng}, K., \& {Nampaisarn}, T. 2011, \apj, 736, 47, \dodoi{10.1088/0004-637X/736/1/47}

\bibitem[{{Burrows} {et~al.}(2001){Burrows}, {Hubbard}, {Lunine}, \& {Liebert}}]{2001RvMP...73..719B}
{Burrows}, A., {Hubbard}, W.~B., {Lunine}, J.~I., \& {Liebert}, J. 2001, Reviews of Modern Physics, 73, 719, \dodoi{10.1103/RevModPhys.73.719}

\bibitem[{{Carmichael}(2023)}]{2023MNRAS.519.5177C}
{Carmichael}, T.~W. 2023, \mnras, 519, 5177, \dodoi{10.1093/mnras/stac3720}

\bibitem[{{Carmichael} {et~al.}(2021){Carmichael}, {Quinn}, {Zhou}, {Grieves}, {Irwin}, {Stassun}, {Vanderburg}, {Winn}, {Bouchy}, {Brasseur}, {Brice{\~n}o}, {Caldwell}, {Charbonneau}, {Collins}, {Colon}, {Eastman}, {Fausnaugh}, {Fong}, {F{\H{u}}r{\'e}sz}, {Huang}, {Jenkins}, {Kielkopf}, {Latham}, {Law}, {Lund}, {Mann}, {Ricker}, {Rodriguez}, {Schwarz}, {Shporer}, {Tenenbaum}, {Wood}, \& {Ziegler}}]{2021AJ....161...97C}
{Carmichael}, T.~W., {Quinn}, S.~N., {Zhou}, G., {et~al.} 2021, \aj, 161, 97, \dodoi{10.3847/1538-3881/abd4e1}

\bibitem[{{Carmichael} {et~al.}(2022){Carmichael}, {Irwin}, {Murgas}, {Pall{\'e}}, {Stassun}, {Bartnik}, {Collins}, {de Leon}, {Esparza-Borges}, {Fedewa}, {Fong}, {Fukui}, {Jenkins}, {Kagetani}, {Latham}, {Lund}, {Mann}, {Moldovan}, {Morgan}, {Narita}, {Painter}, {Parviainen}, {Quintana}, {Ricker}, {Schulte}, {Schwarz}, {Seager}, {Sokolovsky}, {Twicken}, \& {Winn}}]{2022MNRAS.514.4944C}
{Carmichael}, T.~W., {Irwin}, J.~M., {Murgas}, F., {et~al.} 2022, \mnras, 514, 4944, \dodoi{10.1093/mnras/stac1666}

\bibitem[{{Cenarro} {et~al.}(2019){Cenarro}, {Moles}, {Crist{\'o}bal-Hornillos}, {Mar{\'\i}n-Franch}, {Ederoclite}, {Varela}, {L{\'o}pez-Sanjuan}, {Hern{\'a}ndez-Monteagudo}, {Angulo}, {V{\'a}zquez Rami{\'o}}, {Viironen}, {Bonoli}, {Orsi}, {Hurier}, {San Roman}, {Greisel}, {Vilella-Rojo}, {D{\'\i}az-Garc{\'\i}a}, {Logro{\~n}o-Garc{\'\i}a}, {Gurung-L{\'o}pez}, {Spinoso}, {Izquierdo-Villalba}, {Aguerri}, {Allende Prieto}, {Bonatto}, {Carvano}, {Chies-Santos}, {Daflon}, {Dupke}, {Falc{\'o}n-Barroso}, {Gon{\c{c}}alves}, {Jim{\'e}nez-Teja}, {Molino}, {Placco}, {Solano}, {Whitten}, {Abril}, {Ant{\'o}n}, {Bello}, {Bielsa de Toledo}, {Castillo-Ram{\'\i}rez}, {Chueca}, {Civera}, {D{\'\i}az-Mart{\'\i}n}, {Dom{\'\i}nguez-Mart{\'\i}nez}, {Garzar{\'a}n-Calderaro}, {Hern{\'a}ndez-Fuertes}, {Iglesias-Marzoa}, {I{\~n}iguez}, {Jim{\'e}nez Ruiz}, {Kruuse}, {Lamadrid}, {Lasso-Cabrera}, {L{\'o}pez-Alegre}, {L{\'o}pez-Sainz}, {Ma{\'\i}cas}, {Moreno-Signes}, {Muniesa}, {Rodr{\'\i}guez-Llano}, {Rueda-Teruel}, {Rueda-Teruel},
  {Soriano-Lagu{\'\i}a}, {Tilve}, {Valdivielso}, {Yanes-D{\'\i}az}, {Alcaniz}, {Mendes de Oliveira}, {Sodr{\'e}}, {Coelho}, {Lopes de Oliveira}, {Tamm}, {Xavier}, {Abramo}, {Akras}, {Alfaro}, {Alvarez-Candal}, {Ascaso}, {Beasley}, {Beers}, {Borges Fernandes}, {Bruzual}, {Buzzo}, {Carrasco}, {Cepa}, {Cortesi}, {Costa-Duarte}, {De Pr{\'a}}, {Favole}, {Galarza}, {Galbany}, {Garcia}, {Gonz{\'a}lez Delgado}, {Gonz{\'a}lez-Serrano}, {Guti{\'e}rrez-Soto}, {Hernandez-Jimenez}, {Kanaan}, {Kuncarayakti}, {Landim}, {Laur}, {Licandro}, {Lima Neto}, {Lyman}, {Ma{\'\i}z Apell{\'a}niz}, {Miralda-Escud{\'e}}, {Morate}, {Nogueira-Cavalcante}, {Novais}, {Oncins}, {Oteo}, {Overzier}, {Pereira}, {Rebassa-Mansergas}, {Reis}, {Roig}, {Sako}, {Salvador-Rusi{\~n}ol}, {Sampedro}, {S{\'a}nchez-Bl{\'a}zquez}, {Santos}, {Schmidtobreick}, {Siffert}, {Telles}, \& {Vilchez}}]{2019A&A...622A.176C}
{Cenarro}, A.~J., {Moles}, M., {Crist{\'o}bal-Hornillos}, D., {et~al.} 2019, \aap, 622, A176, \dodoi{10.1051/0004-6361/201833036}

\bibitem[{{Chabrier} {et~al.}(2023){Chabrier}, {Baraffe}, {Phillips}, \& {Debras}}]{2023A&A...671A.119C}
{Chabrier}, G., {Baraffe}, I., {Phillips}, M., \& {Debras}, F. 2023, \aap, 671, A119, \dodoi{10.1051/0004-6361/202243832}

\bibitem[{{Clarke} \& {Pringle}(1991)}]{1991MNRAS.249..584C}
{Clarke}, C.~J., \& {Pringle}, J.~E. 1991, \mnras, 249, 584, \dodoi{10.1093/mnras/249.4.584}

\bibitem[{{Dewberry}(2024)}]{2024arXiv240306979D}
{Dewberry}, J.~W. 2024, arXiv e-prints, arXiv:2403.06979, \dodoi{10.48550/arXiv.2403.06979}

\bibitem[{{Dong} {et~al.}(2023){Dong}, {Wang}, {Rice}, {Zhou}, {Huang}, {Dawson}, {Stef{\'a}nsson}, {Halverson}, {Kanodia}, {Mahadevan}, {McElwain}, {Alvarado-Montes}, {Ninan}, {Robertson}, {Roy}, {Schwab}, {Logsdon}, {Terrien}, {Collins}, {Srdoc}, {Sefako}, {Laloum}, {Latham}, {Bieryla}, {Dalba}, {Dragomir}, {Villanueva}, {Howell}, {Ricker}, {Seager}, {Winn}, {Jenkins}, {Shporer}, \& {Rapetti}}]{2023ApJ...951L..29D}
{Dong}, J., {Wang}, S., {Rice}, M., {et~al.} 2023, \apjl, 951, L29, \dodoi{10.3847/2041-8213/acd93d}

\bibitem[{{Duch{\^e}ne} \& {Kraus}(2013)}]{2013ARA&A..51..269D}
{Duch{\^e}ne}, G., \& {Kraus}, A. 2013, \araa, 51, 269, \dodoi{10.1146/annurev-astro-081710-102602}

\bibitem[{{El-Badry} {et~al.}(2021){El-Badry}, {Rix}, \& {Heintz}}]{2021MNRAS.506.2269E}
{El-Badry}, K., {Rix}, H.-W., \& {Heintz}, T.~M. 2021, \mnras, 506, 2269, \dodoi{10.1093/mnras/stab323}

\bibitem[{{Espinoza} \& {Jord{\'a}n}(2016)}]{2016MNRAS.457.3573E}
{Espinoza}, N., \& {Jord{\'a}n}, A. 2016, \mnras, 457, 3573, \dodoi{10.1093/mnras/stw224}

\bibitem[{{Fabian} {et~al.}(1975){Fabian}, {Pringle}, \& {Rees}}]{1975MNRAS.172P..15F}
{Fabian}, A.~C., {Pringle}, J.~E., \& {Rees}, M.~J. 1975, \mnras, 172, 15, \dodoi{10.1093/mnras/172.1.15P}

\bibitem[{{Forbes} \& {Loeb}(2019)}]{2019ApJ...871..227F}
{Forbes}, J.~C., \& {Loeb}, A. 2019, \apj, 871, 227, \dodoi{10.3847/1538-4357/aafac8}

\bibitem[{{Gaia Collaboration} {et~al.}(2023){Gaia Collaboration}, {Vallenari}, {Brown}, {Prusti}, {de Bruijne}, {Arenou}, {Babusiaux}, {Biermann}, {Creevey}, {Ducourant}, {Evans}, {Eyer}, {Guerra}, {Hutton}, {Jordi}, {Klioner}, {Lammers}, {Lindegren}, {Luri}, {Mignard}, {Panem}, {Pourbaix}, {Randich}, {Sartoretti}, {Soubiran}, {Tanga}, {Walton}, {Bailer-Jones}, {Bastian}, {Drimmel}, {Jansen}, {Katz}, {Lattanzi}, {van Leeuwen}, {Bakker}, {Cacciari}, {Casta{\~n}eda}, {De Angeli}, {Fabricius}, {Fouesneau}, {Fr{\'e}mat}, {Galluccio}, {Guerrier}, {Heiter}, {Masana}, {Messineo}, {Mowlavi}, {Nicolas}, {Nienartowicz}, {Pailler}, {Panuzzo}, {Riclet}, {Roux}, {Seabroke}, {Sordo}, {Th{\'e}venin}, {Gracia-Abril}, {Portell}, {Teyssier}, {Altmann}, {Andrae}, {Audard}, {Bellas-Velidis}, {Benson}, {Berthier}, {Blomme}, {Burgess}, {Busonero}, {Busso}, {C{\'a}novas}, {Carry}, {Cellino}, {Cheek}, {Clementini}, {Damerdji}, {Davidson}, {de Teodoro}, {Nu{\~n}ez Campos}, {Delchambre}, {Dell'Oro}, {Esquej},
  {Fern{\'a}ndez-Hern{\'a}ndez}, {Fraile}, {Garabato}, {Garc{\'\i}a-Lario}, {Gosset}, {Haigron}, {Halbwachs}, {Hambly}, {Harrison}, {Hern{\'a}ndez}, {Hestroffer}, {Hodgkin}, {Holl}, {Jan{\ss}en}, {Jevardat de Fombelle}, {Jordan}, {Krone-Martins}, {Lanzafame}, {L{\"o}ffler}, {Marchal}, {Marrese}, {Moitinho}, {Muinonen}, {Osborne}, {Pancino}, {Pauwels}, {Recio-Blanco}, {Reyl{\'e}}, {Riello}, {Rimoldini}, {Roegiers}, {Rybizki}, {Sarro}, {Siopis}, {Smith}, {Sozzetti}, {Utrilla}, {van Leeuwen}, {Abbas}, {{\'A}brah{\'a}m}, {Abreu Aramburu}, {Aerts}, {Aguado}, {Ajaj}, {Aldea-Montero}, {Altavilla}, {{\'A}lvarez}, {Alves}, {Anders}, {Anderson}, {Anglada Varela}, {Antoja}, {Baines}, {Baker}, {Balaguer-N{\'u}{\~n}ez}, {Balbinot}, {Balog}, {Barache}, {Barbato}, {Barros}, {Barstow}, {Bartolom{\'e}}, {Bassilana}, {Bauchet}, {Becciani}, {Bellazzini}, {Berihuete}, {Bernet}, {Bertone}, {Bianchi}, {Binnenfeld}, {Blanco-Cuaresma}, {Blazere}, {Boch}, {Bombrun}, {Bossini}, {Bouquillon}, {Bragaglia}, {Bramante}, {Breedt},
  {Bressan}, {Brouillet}, {Brugaletta}, {Bucciarelli}, {Burlacu}, {Butkevich}, {Buzzi}, {Caffau}, {Cancelliere}, {Cantat-Gaudin}, {Carballo}, {Carlucci}, {Carnerero}, {Carrasco}, {Casamiquela}, {Castellani}, {Castro-Ginard}, {Chaoul}, {Charlot}, {Chemin}, {Chiaramida}, {Chiavassa}, {Chornay}, {Comoretto}, {Contursi}, {Cooper}, {Cornez}, {Cowell}, {Crifo}, {Cropper}, {Crosta}, {Crowley}, {Dafonte}, {Dapergolas}, {David}, {David}, {de Laverny}, {De Luise}, {De March}, {De Ridder}, {de Souza}, {de Torres}, {del Peloso}, {del Pozo}, {Delbo}, {Delgado}, {Delisle}, {Demouchy}, {Dharmawardena}, {Di Matteo}, {Diakite}, {Diener}, {Distefano}, {Dolding}, {Edvardsson}, {Enke}, {Fabre}, {Fabrizio}, {Faigler}, {Fedorets}, {Fernique}, {Fienga}, {Figueras}, {Fournier}, {Fouron}, {Fragkoudi}, {Gai}, {Garcia-Gutierrez}, {Garcia-Reinaldos}, {Garc{\'\i}a-Torres}, {Garofalo}, {Gavel}, {Gavras}, {Gerlach}, {Geyer}, {Giacobbe}, {Gilmore}, {Girona}, {Giuffrida}, {Gomel}, {Gomez}, {Gonz{\'a}lez-N{\'u}{\~n}ez},
  {Gonz{\'a}lez-Santamar{\'\i}a}, {Gonz{\'a}lez-Vidal}, {Granvik}, {Guillout}, {Guiraud}, {Guti{\'e}rrez-S{\'a}nchez}, {Guy}, {Hatzidimitriou}, {Hauser}, {Haywood}, {Helmer}, {Helmi}, {Sarmiento}, {Hidalgo}, {Hilger}, {H{\l}adczuk}, {Hobbs}, {Holland}, {Huckle}, {Jardine}, {Jasniewicz}, {Jean-Antoine Piccolo}, {Jim{\'e}nez-Arranz}, {Jorissen}, {Juaristi Campillo}, {Julbe}, {Karbevska}, {Kervella}, {Khanna}, {Kontizas}, {Kordopatis}, {Korn}, {K{\'o}sp{\'a}l}, {Kostrzewa-Rutkowska}, {Kruszy{\'n}ska}, {Kun}, {Laizeau}, {Lambert}, {Lanza}, {Lasne}, {Le Campion}, {Lebreton}, {Lebzelter}, {Leccia}, {Leclerc}, {Lecoeur-Taibi}, {Liao}, {Licata}, {Lindstr{\o}m}, {Lister}, {Livanou}, {Lobel}, {Lorca}, {Loup}, {Madrero Pardo}, {Magdaleno Romeo}, {Managau}, {Mann}, {Manteiga}, {Marchant}, {Marconi}, {Marcos}, {Marcos Santos}, {Mar{\'\i}n Pina}, {Marinoni}, {Marocco}, {Marshall}, {Martin Polo}, {Mart{\'\i}n-Fleitas}, {Marton}, {Mary}, {Masip}, {Massari}, {Mastrobuono-Battisti}, {Mazeh}, {McMillan}, {Messina}, {Michalik},
  {Millar}, {Mints}, {Molina}, {Molinaro}, {Moln{\'a}r}, {Monari}, {Mongui{\'o}}, {Montegriffo}, {Montero}, {Mor}, {Mora}, {Morbidelli}, {Morel}, {Morris}, {Muraveva}, {Murphy}, {Musella}, {Nagy}, {Noval}, {Oca{\~n}a}, {Ogden}, {Ordenovic}, {Osinde}, {Pagani}, {Pagano}, {Palaversa}, {Palicio}, {Pallas-Quintela}, {Panahi}, {Payne-Wardenaar}, {Pe{\~n}alosa Esteller}, {Penttil{\"a}}, {Pichon}, {Piersimoni}, {Pineau}, {Plachy}, {Plum}, {Poggio}, {Pr{\v{s}}a}, {Pulone}, {Racero}, {Ragaini}, {Rainer}, {Raiteri}, {Rambaux}, {Ramos}, {Ramos-Lerate}, {Re Fiorentin}, {Regibo}, {Richards}, {Rios Diaz}, {Ripepi}, {Riva}, {Rix}, {Rixon}, {Robichon}, {Robin}, {Robin}, {Roelens}, {Rogues}, {Rohrbasser}, {Romero-G{\'o}mez}, {Rowell}, {Royer}, {Ruz Mieres}, {Rybicki}, {Sadowski}, {S{\'a}ez N{\'u}{\~n}ez}, {Sagrist{\`a} Sell{\'e}s}, {Sahlmann}, {Salguero}, {Samaras}, {Sanchez Gimenez}, {Sanna}, {Santove{\~n}a}, {Sarasso}, {Schultheis}, {Sciacca}, {Segol}, {Segovia}, {S{\'e}gransan}, {Semeux}, {Shahaf}, {Siddiqui}, {Siebert},
  {Siltala}, {Silvelo}, {Slezak}, {Slezak}, {Smart}, {Snaith}, {Solano}, {Solitro}, {Souami}, {Souchay}, {Spagna}, {Spina}, {Spoto}, {Steele}, {Steidelm{\"u}ller}, {Stephenson}, {S{\"u}veges}, {Surdej}, {Szabados}, {Szegedi-Elek}, {Taris}, {Taylor}, {Teixeira}, {Tolomei}, {Tonello}, {Torra}, {Torra}, {Torralba Elipe}, {Trabucchi}, {Tsounis}, {Turon}, {Ulla}, {Unger}, {Vaillant}, {van Dillen}, {van Reeven}, {Vanel}, {Vecchiato}, {Viala}, {Vicente}, {Voutsinas}, {Weiler}, {Wevers}, {Wyrzykowski}, {Yoldas}, {Yvard}, {Zhao}, {Zorec}, {Zucker}, \& {Zwitter}}]{2023A&A...674A...1G}
{Gaia Collaboration}, {Vallenari}, A., {Brown}, A.~G.~A., {et~al.} 2023, \aap, 674, A1, \dodoi{10.1051/0004-6361/202243940}

\bibitem[{{Gaudi}(2003)}]{2003astro.ph..7280G}
{Gaudi}, B.~S. 2003, arXiv e-prints, astro, \dodoi{10.48550/arXiv.astro-ph/0307280}

\bibitem[{{Gilliland} {et~al.}(2000){Gilliland}, {Brown}, {Guhathakurta}, {Sarajedini}, {Milone}, {Albrow}, {Baliber}, {Bruntt}, {Burrows}, {Charbonneau}, {Choi}, {Cochran}, {Edmonds}, {Frandsen}, {Howell}, {Lin}, {Marcy}, {Mayor}, {Naef}, {Sigurdsson}, {Stagg}, {Vandenberg}, {Vogt}, \& {Williams}}]{gilliland2000lack}
{Gilliland}, R.~L., {Brown}, T.~M., {Guhathakurta}, P., {et~al.} 2000, \apjl, 545, L47, \dodoi{10.1086/317334}

\bibitem[{{Goodwin} \& {Whitworth}(2007)}]{2007A&A...466..943G}
{Goodwin}, S.~P., \& {Whitworth}, A. 2007, \aap, 466, 943, \dodoi{10.1051/0004-6361:20066745}

\bibitem[{{Goodwin} {et~al.}(2004){Goodwin}, {Whitworth}, \& {Ward-Thompson}}]{2004A&A...414..633G}
{Goodwin}, S.~P., {Whitworth}, A.~P., \& {Ward-Thompson}, D. 2004, \aap, 414, 633, \dodoi{10.1051/0004-6361:20031594}

\bibitem[{{G{\"u}nther} \& {Daylan}(2021)}]{2021ApJS..254...13G}
{G{\"u}nther}, M.~N., \& {Daylan}, T. 2021, \apjs, 254, 13, \dodoi{10.3847/1538-4365/abe70e}

\bibitem[{{Hale}(1994)}]{1994AJ....107..306H}
{Hale}, A. 1994, \aj, 107, 306, \dodoi{10.1086/116855}

\bibitem[{{Harris} {et~al.}(2020){Harris}, {Millman}, {van der Walt}, {Gommers}, {Virtanen}, {Cournapeau}, {Wieser}, {Taylor}, {Berg}, {Smith}, {Kern}, {Picus}, {Hoyer}, {van Kerkwijk}, {Brett}, {Haldane}, {del R{\'\i}o}, {Wiebe}, {Peterson}, {G{\'e}rard-Marchant}, {Sheppard}, {Reddy}, {Weckesser}, {Abbasi}, {Gohlke}, \& {Oliphant}}]{2020Natur.585..357H}
{Harris}, C.~R., {Millman}, K.~J., {van der Walt}, S.~J., {et~al.} 2020, \nat, 585, 357, \dodoi{10.1038/s41586-020-2649-2}

\bibitem[{{Herpich} {et~al.}(2021){Herpich}, {Ferreira Lopes}, {Saito}, {Minniti}, {Ederoclite}, {Ferreira}, \& {Catelan}}]{2021A&A...647A.169H}
{Herpich}, F.~R., {Ferreira Lopes}, C.~E., {Saito}, R.~K., {et~al.} 2021, \aap, 647, A169, \dodoi{10.1051/0004-6361/201834356}

\bibitem[{{Hixenbaugh} {et~al.}(2023){Hixenbaugh}, {Wang}, {Rice}, \& {Wang}}]{2023ApJ...949L..35H}
{Hixenbaugh}, K., {Wang}, X.-Y., {Rice}, M., \& {Wang}, S. 2023, \apjl, 949, L35, \dodoi{10.3847/2041-8213/acd6f5}

\bibitem[{{H{\o}g} {et~al.}(2000){H{\o}g}, {Fabricius}, {Makarov}, {Bastian}, {Schwekendiek}, {Wicenec}, {Urban}, {Corbin}, \& {Wycoff}}]{2000A&A...357..367H}
{H{\o}g}, E., {Fabricius}, C., {Makarov}, V.~V., {et~al.} 2000, \aap, 357, 367

\bibitem[{{Holt}(1893)}]{1893AstAp..12..646H}
{Holt}, J.~R. 1893, Astronomy and Astro-Physics (formerly The Sidereal Messenger), 12, 646

\bibitem[{{Hoyle}(1953)}]{1953ApJ...118..513H}
{Hoyle}, F. 1953, \apj, 118, 513, \dodoi{10.1086/145780}

\bibitem[{{Hu} {et~al.}(2024){Hu}, {Rice}, {Wang}, {Wang}, {Shporer}, {Teske}, {Yee}, {Butler}, {Shectman}, {Crane}, {Collins}, \& {Collins}}]{2024AJ....167..175H}
{Hu}, Q., {Rice}, M., {Wang}, X.-Y., {et~al.} 2024, \aj, 167, 175, \dodoi{10.3847/1538-3881/ad2855}

\bibitem[{{Huber} {et~al.}(2009){Huber}, {Stello}, {Bedding}, {Chaplin}, {Arentoft}, {Quirion}, \& {Kjeldsen}}]{2009CoAst.160...74H}
{Huber}, D., {Stello}, D., {Bedding}, T.~R., {et~al.} 2009, Communications in Asteroseismology, 160, 74, \dodoi{10.48550/arXiv.0910.2764}

\bibitem[{{Hunter}(2007)}]{2007CSE.....9...90H}
{Hunter}, J.~D. 2007, Computing in Science and Engineering, 9, 90, \dodoi{10.1109/MCSE.2007.55}

\bibitem[{{Hut}(1981)}]{1981A&A....99..126H}
{Hut}, P. 1981, \aap, 99, 126

\bibitem[{{Ivshina} \& {Winn}(2022)}]{2022ApJS..259...62I}
{Ivshina}, E.~S., \& {Winn}, J.~N. 2022, \apjs, 259, 62, \dodoi{10.3847/1538-4365/ac545b}

\bibitem[{{Jackson} {et~al.}(2008){Jackson}, {Greenberg}, \& {Barnes}}]{2008ApJ...678.1396J}
{Jackson}, B., {Greenberg}, R., \& {Barnes}, R. 2008, \apj, 678, 1396, \dodoi{10.1086/529187}

\bibitem[{{Jenkins} {et~al.}(2016){Jenkins}, {Twicken}, {McCauliff}, {Campbell}, {Sanderfer}, {Lung}, {Mansouri-Samani}, {Girouard}, {Tenenbaum}, {Klaus}, {Smith}, {Caldwell}, {Chacon}, {Henze}, {Heiges}, {Latham}, {Morgan}, {Swade}, {Rinehart}, \& {Vanderspek}}]{2016SPIE.9913E..3EJ}
{Jenkins}, J.~M., {Twicken}, J.~D., {McCauliff}, S., {et~al.} 2016, in Society of Photo-Optical Instrumentation Engineers (SPIE) Conference Series, Vol. 9913, Software and Cyberinfrastructure for Astronomy IV, ed. G.~{Chiozzi} \& J.~C. {Guzman}, 99133E, \dodoi{10.1117/12.2233418}

\bibitem[{{Kamdar} {et~al.}(2019){Kamdar}, {Conroy}, {Ting}, {Bonaca}, {Johnson}, \& {Cargile}}]{2019ApJ...884..173K}
{Kamdar}, H., {Conroy}, C., {Ting}, Y.-S., {et~al.} 2019, \apj, 884, 173, \dodoi{10.3847/1538-4357/ab44be}

\bibitem[{{Kaplan} {et~al.}(2019){Kaplan}, {Bender}, {Terrien}, {Ninan}, {Roy}, \& {Mahadevan}}]{2019ASPC..523..567K}
{Kaplan}, K.~F., {Bender}, C.~F., {Terrien}, R.~C., {et~al.} 2019, in Astronomical Society of the Pacific Conference Series, Vol. 523, Astronomical Data Analysis Software and Systems XXVII, ed. P.~J. {Teuben}, M.~W. {Pound}, B.~A. {Thomas}, \& E.~M. {Warner}, 567

\bibitem[{{Kaplan} {et~al.}(2012){Kaplan}, {Stamatellos}, \& {Whitworth}}]{2012Ap&SS.341..395K}
{Kaplan}, M., {Stamatellos}, D., \& {Whitworth}, A.~P. 2012, \apss, 341, 395, \dodoi{10.1007/s10509-012-1110-x}

\bibitem[{{Kesseli} {et~al.}(2017){Kesseli}, {West}, {Veyette}, {Harrison}, {Feldman}, \& {Bochanski}}]{2017ApJS..230...16K}
{Kesseli}, A.~Y., {West}, A.~A., {Veyette}, M., {et~al.} 2017, \apjs, 230, 16, \dodoi{10.3847/1538-4365/aa656d}

\bibitem[{{Kiefer} {et~al.}(2019){Kiefer}, {H{\'e}brard}, {Sahlmann}, {Sousa}, {Forveille}, {Santos}, {Mayor}, {Deleuil}, {Wilson}, {Dalal}, {D{\'\i}az}, {Henry}, {Hagelberg}, {Hobson}, {Demangeon}, {Bourrier}, {Delfosse}, {Arnold}, {Astudillo-Defru}, {Beuzit}, {Boisse}, {Bonfils}, {Borgniet}, {Bouchy}, {Courcol}, {Ehrenreich}, {Hara}, {Lagrange}, {Lovis}, {Montagnier}, {Moutou}, {Pepe}, {Perrier}, {Rey}, {Santerne}, {S{\'e}gransan}, {Udry}, \& {Vidal-Madjar}}]{2019A&A...631A.125K}
{Kiefer}, F., {H{\'e}brard}, G., {Sahlmann}, J., {et~al.} 2019, \aap, 631, A125, \dodoi{10.1051/0004-6361/201935113}

\bibitem[{{Kippenhahn} {et~al.}(2013){Kippenhahn}, {Weigert}, \& {Weiss}}]{2013sse..book.....K}
{Kippenhahn}, R., {Weigert}, A., \& {Weiss}, A. 2013, {Stellar Structure and Evolution}, \dodoi{10.1007/978-3-642-30304-3}

\bibitem[{{Kipping}(2013)}]{2013MNRAS.435.2152K}
{Kipping}, D.~M. 2013, \mnras, 435, 2152, \dodoi{10.1093/mnras/stt1435}

\bibitem[{{Kirkpatrick} {et~al.}(2008){Kirkpatrick}, {Cruz}, {Barman}, {Burgasser}, {Looper}, {Tinney}, {Gelino}, {Lowrance}, {Liebert}, {Carpenter}, {Hillenbrand}, \& {Stauffer}}]{2008ApJ...689.1295K}
{Kirkpatrick}, J.~D., {Cruz}, K.~L., {Barman}, T.~S., {et~al.} 2008, \apj, 689, 1295, \dodoi{10.1086/592768}

\bibitem[{{Kraft}(1967)}]{1967ApJ...150..551K}
{Kraft}, R.~P. 1967, \apj, 150, 551, \dodoi{10.1086/149359}

\bibitem[{{Kratter}(2011)}]{2011ASPC..447...47K}
{Kratter}, K.~M. 2011, in Astronomical Society of the Pacific Conference Series, Vol. 447, Evolution of Compact Binaries, ed. L.~{Schmidtobreick}, M.~R. {Schreiber}, \& C.~{Tappert}, 47, \dodoi{10.48550/arXiv.1109.3740}

\bibitem[{{Kratter} {et~al.}(2010){Kratter}, {Matzner}, {Krumholz}, \& {Klein}}]{2010ApJ...708.1585K}
{Kratter}, K.~M., {Matzner}, C.~D., {Krumholz}, M.~R., \& {Klein}, R.~I. 2010, \apj, 708, 1585, \dodoi{10.1088/0004-637X/708/2/1585}

\bibitem[{{Krumholz} {et~al.}(2009){Krumholz}, {Klein}, {McKee}, {Offner}, \& {Cunningham}}]{2009Sci...323..754K}
{Krumholz}, M.~R., {Klein}, R.~I., {McKee}, C.~F., {Offner}, S. S.~R., \& {Cunningham}, A.~J. 2009, Science, 323, 754, \dodoi{10.1126/science.1165857}

\bibitem[{{Kuffmeier} {et~al.}(2019){Kuffmeier}, {Calcutt}, \& {Kristensen}}]{2019A&A...628A.112K}
{Kuffmeier}, M., {Calcutt}, H., \& {Kristensen}, L.~E. 2019, \aap, 628, A112, \dodoi{10.1051/0004-6361/201935504}

\bibitem[{{Kumar}(1963{\natexlab{a}})}]{1963ApJ...137.1121K}
{Kumar}, S.~S. 1963{\natexlab{a}}, \apj, 137, 1121, \dodoi{10.1086/147589}

\bibitem[{{Kumar}(1963{\natexlab{b}})}]{1963ApJ...137.1126K}
---. 1963{\natexlab{b}}, \apj, 137, 1126, \dodoi{10.1086/147590}

\bibitem[{{Kunimoto} {et~al.}(2022){Kunimoto}, {Tey}, {Fong}, {Hesse}, {Shporer}, {Fausnaugh}, {Vanderspek}, \& {Ricker}}]{2022RNAAS...6..236K}
{Kunimoto}, M., {Tey}, E., {Fong}, W., {et~al.} 2022, Research Notes of the American Astronomical Society, 6, 236, \dodoi{10.3847/2515-5172/aca158}

\bibitem[{{Larson}(1969)}]{1969MNRAS.145..271L}
{Larson}, R.~B. 1969, \mnras, 145, 271, \dodoi{10.1093/mnras/145.3.271}

\bibitem[{{Laughlin} \& {Bodenheimer}(1994)}]{1994ApJ...436..335L}
{Laughlin}, G., \& {Bodenheimer}, P. 1994, \apj, 436, 335, \dodoi{10.1086/174909}

\bibitem[{{Lee} {et~al.}(2017){Lee}, {Lee}, {Dunham}, {Tatematsu}, {Choi}, {Bergin}, \& {Evans}}]{2017NatAs...1E.172L}
{Lee}, J.-E., {Lee}, S., {Dunham}, M.~M., {et~al.} 2017, Nature Astronomy, 1, 0172, \dodoi{10.1038/s41550-017-0172}

\bibitem[{Li(2023)}]{jiaxuan_li_2023_8126529}
Li, J. 2023, AstroJacobLi/smplotlib: v0.0.9, v0.0.9,  Zenodo, \dodoi{10.5281/zenodo.8126529}

\bibitem[{{Lightkurve Collaboration} {et~al.}(2018){Lightkurve Collaboration}, {Cardoso}, {Hedges}, {Gully-Santiago}, {Saunders}, {Cody}, {Barclay}, {Hall}, {Sagear}, {Turtelboom}, {Zhang}, {Tzanidakis}, {Mighell}, {Coughlin}, {Bell}, {Berta-Thompson}, {Williams}, {Dotson}, \& {Barentsen}}]{2018ascl.soft12013L}
{Lightkurve Collaboration}, {Cardoso}, J. V. d.~M., {Hedges}, C., {et~al.} 2018, {Lightkurve: Kepler and TESS time series analysis in Python}, Astrophysics Source Code Library, record ascl:1812.013.
\newblock \doeprint{1812.013}

\bibitem[{{Louden} {et~al.}(2021){Louden}, {Winn}, {Petigura}, {Isaacson}, {Howard}, {Masuda}, {Albrecht}, \& {Kosiarek}}]{2021AJ....161...68L}
{Louden}, E.~M., {Winn}, J.~N., {Petigura}, E.~A., {et~al.} 2021, \aj, 161, 68, \dodoi{10.3847/1538-3881/abcebd}

\bibitem[{{Lubin} {et~al.}(2023){Lubin}, {Wang}, {Rice}, {Dong}, {Wang}, {Radzom}, {Robertson}, {Stefansson}, {Alvarado-Montes}, {Beard}, {Bender}, {Gupta}, {Halverson}, {Kanodia}, {Li}, {Lin}, {Logsdon}, {Lubar}, {Mahadevan}, {Ninan}, {Rajagopal}, {Roy}, {Schwab}, \& {Wright}}]{2023ApJ...959L...5L}
{Lubin}, J., {Wang}, X.-Y., {Rice}, M., {et~al.} 2023, \apjl, 959, L5, \dodoi{10.3847/2041-8213/ad0fea}

\bibitem[{{Ma} \& {Ge}(2014)}]{2014MNRAS.439.2781M}
{Ma}, B., \& {Ge}, J. 2014, \mnras, 439, 2781, \dodoi{10.1093/mnras/stu134}

\bibitem[{{Maldonado} \& {Villaver}(2017)}]{2017A&A...602A..38M}
{Maldonado}, J., \& {Villaver}, E. 2017, \aap, 602, A38, \dodoi{10.1051/0004-6361/201630120}

\bibitem[{{Mart{\'\i}n} {et~al.}(2022){Mart{\'\i}n}, {Lodieu}, \& {del Burgo}}]{2022MNRAS.510.2841M}
{Mart{\'\i}n}, E.~L., {Lodieu}, N., \& {del Burgo}, C. 2022, \mnras, 510, 2841, \dodoi{10.1093/mnras/stab2969}

\bibitem[{{Matsumura} {et~al.}(2010){Matsumura}, {Peale}, \& {Rasio}}]{2010ApJ...725.1995M}
{Matsumura}, S., {Peale}, S.~J., \& {Rasio}, F.~A. 2010, \apj, 725, 1995, \dodoi{10.1088/0004-637X/725/2/1995}

\bibitem[{{McLaughlin}(1924)}]{1924ApJ....60...22M}
{McLaughlin}, D.~B. 1924, \apj, 60, 22, \dodoi{10.1086/142826}

\bibitem[{{Mink}(2011)}]{2011ASPC..442..305M}
{Mink}, D.~J. 2011, in Astronomical Society of the Pacific Conference Series, Vol. 442, Astronomical Data Analysis Software and Systems XX, ed. I.~N. {Evans}, A.~{Accomazzi}, D.~J. {Mink}, \& A.~H. {Rots}, 305

\bibitem[{{Naoz} {et~al.}(2011){Naoz}, {Farr}, {Lithwick}, {Rasio}, \& {Teyssandier}}]{2011Natur.473..187N}
{Naoz}, S., {Farr}, W.~M., {Lithwick}, Y., {Rasio}, F.~A., \& {Teyssandier}, J. 2011, \nat, 473, 187, \dodoi{10.1038/nature10076}

\bibitem[{{Naoz} {et~al.}(2012){Naoz}, {Farr}, \& {Rasio}}]{2012ApJ...754L..36N}
{Naoz}, S., {Farr}, W.~M., \& {Rasio}, F.~A. 2012, \apjl, 754, L36, \dodoi{10.1088/2041-8205/754/2/L36}

\bibitem[{{Offner} {et~al.}(2016){Offner}, {Dunham}, {Lee}, {Arce}, \& {Fielding}}]{2016ApJ...827L..11O}
{Offner}, S. S.~R., {Dunham}, M.~M., {Lee}, K.~I., {Arce}, H.~G., \& {Fielding}, D.~B. 2016, \apjl, 827, L11, \dodoi{10.3847/2041-8205/827/1/L11}

\bibitem[{{Offner} {et~al.}(2010){Offner}, {Kratter}, {Matzner}, {Krumholz}, \& {Klein}}]{2010ApJ...725.1485O}
{Offner}, S. S.~R., {Kratter}, K.~M., {Matzner}, C.~D., {Krumholz}, M.~R., \& {Klein}, R.~I. 2010, \apj, 725, 1485, \dodoi{10.1088/0004-637X/725/2/1485}

\bibitem[{{Offner} {et~al.}(2023){Offner}, {Moe}, {Kratter}, {Sadavoy}, {Jensen}, \& {Tobin}}]{2023ASPC..534..275O}
{Offner}, S.~S.~R., {Moe}, M., {Kratter}, K.~M., {et~al.} 2023, in Astronomical Society of the Pacific Conference Series, Vol. 534, Protostars and Planets VII, ed. S.~{Inutsuka}, Y.~{Aikawa}, T.~{Muto}, K.~{Tomida}, \& M.~{Tamura}, 275, \dodoi{10.48550/arXiv.2203.10066}

\bibitem[{{Padoan} \& {Nordlund}(2002)}]{2002ApJ...576..870P}
{Padoan}, P., \& {Nordlund}, {\r{A}}. 2002, \apj, 576, 870, \dodoi{10.1086/341790}

\bibitem[{{Perets} \& {Kratter}(2012)}]{2012ApJ...760...99P}
{Perets}, H.~B., \& {Kratter}, K.~M. 2012, \apj, 760, 99, \dodoi{10.1088/0004-637X/760/2/99}

\bibitem[{{Petrovich}(2015)}]{2015ApJ...805...75P}
{Petrovich}, C. 2015, \apj, 805, 75, \dodoi{10.1088/0004-637X/805/1/75}

\bibitem[{{Prialnik}(2009)}]{2009itss.book.....P}
{Prialnik}, D. 2009, {An Introduction to the Theory of Stellar Structure and Evolution}

\bibitem[{{Psaridi} {et~al.}(2022){Psaridi}, {Bouchy}, {Lendl}, {Grieves}, {Stassun}, {Carmichael}, {Gill}, {Pe{\~n}a Rojas}, {Gan}, {Shporer}, {Bieryla}, {Brahm}, {Christiansen}, {Crossfield}, {Galland}, {Hooton}, {Jenkins}, {Jenkins}, {Latham}, {Lund}, {Rodriguez}, {Ting}, {Udry}, {Ulmer-Moll}, {Wittenmyer}, {Zhang}, {Zhou}, {Addison}, {Cointepas}, {Collins}, {Collins}, {Deline}, {Dressing}, {Evans}, {Giacalone}, {Heitzmann}, {Mireles}, {Mounzer}, {Otegi}, {Radford}, {Rudat}, {Schlieder}, {Schwarz}, {Srdoc}, {Stockdale}, {Suarez}, {Wright}, \& {Zhao}}]{2022A&A...664A..94P}
{Psaridi}, A., {Bouchy}, F., {Lendl}, M., {et~al.} 2022, \aap, 664, A94, \dodoi{10.1051/0004-6361/202243454}

\bibitem[{{Radzom} {et~al.}(2024){Radzom}, {Dong}, {Rice}, {Wang}, {Yee}, {Fairnington}, {Petrovich}, \& {Wang}}]{2024arXiv240406504R}
{Radzom}, B.~T., {Dong}, J., {Rice}, M., {et~al.} 2024, arXiv e-prints, arXiv:2404.06504, \dodoi{10.48550/arXiv.2404.06504}

\bibitem[{{Rice} {et~al.}(2024){Rice}, {Gerbig}, \& {Vanderburg}}]{2024AJ....167..126R}
{Rice}, M., {Gerbig}, K., \& {Vanderburg}, A. 2024, \aj, 167, 126, \dodoi{10.3847/1538-3881/ad1bed}

\bibitem[{{Rice} {et~al.}(2023{\natexlab{a}}){Rice}, {Wang}, {Gerbig}, {Wang}, {Dai}, {Tyler}, {Isaacson}, \& {Howard}}]{2023AJ....165...65R}
{Rice}, M., {Wang}, S., {Gerbig}, K., {et~al.} 2023{\natexlab{a}}, \aj, 165, 65, \dodoi{10.3847/1538-3881/aca88e}

\bibitem[{{Rice} {et~al.}(2022{\natexlab{a}}){Rice}, {Wang}, \& {Laughlin}}]{2022ApJ...926L..17R}
{Rice}, M., {Wang}, S., \& {Laughlin}, G. 2022{\natexlab{a}}, \apjl, 926, L17, \dodoi{10.3847/2041-8213/ac502d}

\bibitem[{{Rice} {et~al.}(2021){Rice}, {Wang}, {Howard}, {Isaacson}, {Dai}, {Wang}, {Beard}, {Behmard}, {Brinkman}, {Rubenzahl}, \& {Laughlin}}]{2021AJ....162..182R}
{Rice}, M., {Wang}, S., {Howard}, A.~W., {et~al.} 2021, \aj, 162, 182, \dodoi{10.3847/1538-3881/ac1f8f}

\bibitem[{{Rice} {et~al.}(2022{\natexlab{b}}){Rice}, {Wang}, {Wang}, {Stef{\'a}nsson}, {Isaacson}, {Howard}, {Logsdon}, {Schweiker}, {Dai}, {Brinkman}, {Giacalone}, \& {Holcomb}}]{2022AJ....164..104R}
{Rice}, M., {Wang}, S., {Wang}, X.-Y., {et~al.} 2022{\natexlab{b}}, \aj, 164, 104, \dodoi{10.3847/1538-3881/ac8153}

\bibitem[{{Rice} {et~al.}(2023{\natexlab{b}}){Rice}, {Wang}, {Wang}, {Shporer}, {Barkaoui}, {Brahm}, {Collins}, {Jord{\'a}n}, {Lowson}, {Butler}, {Crane}, {Shectman}, {Teske}, {Osip}, {Collins}, {Murgas}, {Boyle}, {Pozuelos}, {Timmermans}, {Jehin}, \& {Gillon}}]{2023AJ....166..266R}
{Rice}, M., {Wang}, X.-Y., {Wang}, S., {et~al.} 2023{\natexlab{b}}, \aj, 166, 266, \dodoi{10.3847/1538-3881/ad09de}

\bibitem[{{Ricker} {et~al.}(2015){Ricker}, {Winn}, {Vanderspek}, {Latham}, {Bakos}, {Bean}, {Berta-Thompson}, {Brown}, {Buchhave}, {Butler}, {Butler}, {Chaplin}, {Charbonneau}, {Christensen-Dalsgaard}, {Clampin}, {Deming}, {Doty}, {De Lee}, {Dressing}, {Dunham}, {Endl}, {Fressin}, {Ge}, {Henning}, {Holman}, {Howard}, {Ida}, {Jenkins}, {Jernigan}, {Johnson}, {Kaltenegger}, {Kawai}, {Kjeldsen}, {Laughlin}, {Levine}, {Lin}, {Lissauer}, {MacQueen}, {Marcy}, {McCullough}, {Morton}, {Narita}, {Paegert}, {Palle}, {Pepe}, {Pepper}, {Quirrenbach}, {Rinehart}, {Sasselov}, {Sato}, {Seager}, {Sozzetti}, {Stassun}, {Sullivan}, {Szentgyorgyi}, {Torres}, {Udry}, \& {Villasenor}}]{2015JATIS...1a4003R}
{Ricker}, G.~R., {Winn}, J.~N., {Vanderspek}, R., {et~al.} 2015, Journal of Astronomical Telescopes, Instruments, and Systems, 1, 014003, \dodoi{10.1117/1.JATIS.1.1.014003}

\bibitem[{{Rossiter}(1924)}]{1924ApJ....60...15R}
{Rossiter}, R.~A. 1924, \apj, 60, 15, \dodoi{10.1086/142825}

\bibitem[{{Schlaufman}(2010)}]{2010ApJ...719..602S}
{Schlaufman}, K.~C. 2010, \apj, 719, 602, \dodoi{10.1088/0004-637X/719/1/602}

\bibitem[{{Schlaufman}(2018)}]{2018ApJ...853...37S}
---. 2018, \apj, 853, 37, \dodoi{10.3847/1538-4357/aa961c}

\bibitem[{{Schmidt} {et~al.}(2023){Schmidt}, {Schlaufman}, {Ding}, {Grunblatt}, {Carmichael}, {Bieryla}, {Rodriguez}, {Schulte}, {Vowell}, {Zhou}, {Quinn}, {Yee}, {Winn}, {Hartman}, {Latham}, {Caldwell}, {Fausnaugh}, {Hedges}, {Jenkins}, {Osborn}, \& {Seager}}]{2023AJ....166..225S}
{Schmidt}, S.~P., {Schlaufman}, K.~C., {Ding}, K., {et~al.} 2023, \aj, 166, 225, \dodoi{10.3847/1538-3881/ad0135}

\bibitem[{{Schwab} {et~al.}(2016){Schwab}, {Rakich}, {Gong}, {Mahadevan}, {Halverson}, {Roy}, {Terrien}, {Robertson}, {Hearty}, {Levi}, {Monson}, {Wright}, {McElwain}, {Bender}, {Blake}, {St{\"u}rmer}, {Gurevich}, {Chakraborty}, \& {Ramsey}}]{2016SPIE.9908E..7HS}
{Schwab}, C., {Rakich}, A., {Gong}, Q., {et~al.} 2016, in Society of Photo-Optical Instrumentation Engineers (SPIE) Conference Series, Vol. 9908, Ground-based and Airborne Instrumentation for Astronomy VI, ed. C.~J. {Evans}, L.~{Simard}, \& H.~{Takami}, 99087H, \dodoi{10.1117/12.2234411}

\bibitem[{{Siverd} {et~al.}(2012){Siverd}, {Beatty}, {Pepper}, {Eastman}, {Collins}, {Bieryla}, {Latham}, {Buchhave}, {Jensen}, {Crepp}, {Street}, {Stassun}, {Gaudi}, {Berlind}, {Calkins}, {DePoy}, {Esquerdo}, {Fulton}, {F{\H{u}}r{\'e}sz}, {Geary}, {Gould}, {Hebb}, {Kielkopf}, {Marshall}, {Pogge}, {Stanek}, {Stefanik}, {Szentgyorgyi}, {Trueblood}, {Trueblood}, {Stutz}, \& {van Saders}}]{2012ApJ...761..123S}
{Siverd}, R.~J., {Beatty}, T.~G., {Pepper}, J., {et~al.} 2012, \apj, 761, 123, \dodoi{10.1088/0004-637X/761/2/123}

\bibitem[{{Skrutskie} {et~al.}(2006){Skrutskie}, {Cutri}, {Stiening}, {Weinberg}, {Schneider}, {Carpenter}, {Beichman}, {Capps}, {Chester}, {Elias}, {Huchra}, {Liebert}, {Lonsdale}, {Monet}, {Price}, {Seitzer}, {Jarrett}, {Kirkpatrick}, {Gizis}, {Howard}, {Evans}, {Fowler}, {Fullmer}, {Hurt}, {Light}, {Kopan}, {Marsh}, {McCallon}, {Tam}, {Van Dyk}, \& {Wheelock}}]{2006AJ....131.1163S}
{Skrutskie}, M.~F., {Cutri}, R.~M., {Stiening}, R., {et~al.} 2006, \aj, 131, 1163, \dodoi{10.1086/498708}

\bibitem[{{Southworth}(2011)}]{2011MNRAS.417.2166S}
{Southworth}, J. 2011, \mnras, 417, 2166, \dodoi{10.1111/j.1365-2966.2011.19399.x}

\bibitem[{{Spalding} \& {Batygin}(2015)}]{2015ApJ...811...82S}
{Spalding}, C., \& {Batygin}, K. 2015, \apj, 811, 82, \dodoi{10.1088/0004-637X/811/2/82}

\bibitem[{{Spalding} \& {Winn}(2022)}]{2022ApJ...927...22S}
{Spalding}, C., \& {Winn}, J.~N. 2022, \apj, 927, 22, \dodoi{10.3847/1538-4357/ac4993}

\bibitem[{{Speagle}(2020)}]{2020MNRAS.493.3132S}
{Speagle}, J.~S. 2020, \mnras, 493, 3132, \dodoi{10.1093/mnras/staa278}

\bibitem[{{Spiegel} {et~al.}(2011){Spiegel}, {Burrows}, \& {Milsom}}]{2011ApJ...727...57S}
{Spiegel}, D.~S., {Burrows}, A., \& {Milsom}, J.~A. 2011, \apj, 727, 57, \dodoi{10.1088/0004-637X/727/1/57}

\bibitem[{{Teyssandier} {et~al.}(2019){Teyssandier}, {Lai}, \& {Vick}}]{2019MNRAS.486.2265T}
{Teyssandier}, J., {Lai}, D., \& {Vick}, M. 2019, \mnras, 486, 2265, \dodoi{10.1093/mnras/stz1011}

\bibitem[{{The pandas development team}(2023)}]{2022zndo...3509134T}
{The pandas development team}. 2023, {pandas-dev/pandas: Pandas}, v2.2.0rc0,  Zenodo, \dodoi{10.5281/zenodo.3509134}

\bibitem[{{Tohline}(2002)}]{2002ARA&A..40..349T}
{Tohline}, J.~E. 2002, \araa, 40, 349, \dodoi{10.1146/annurev.astro.40.060401.093810}

\bibitem[{{Tokovinin} \& {Moe}(2020)}]{2020MNRAS.491.5158T}
{Tokovinin}, A., \& {Moe}, M. 2020, \mnras, 491, 5158, \dodoi{10.1093/mnras/stz3299}

\bibitem[{{Triaud} {et~al.}(2009){Triaud}, {Queloz}, {Bouchy}, {Moutou}, {Collier Cameron}, {Claret}, {Barge}, {Benz}, {Deleuil}, {Guillot}, {H{\'e}brard}, {Lecavelier Des {\'E}tangs}, {Lovis}, {Mayor}, {Pepe}, \& {Udry}}]{2009A&A...506..377T}
{Triaud}, A.~H.~M.~J., {Queloz}, D., {Bouchy}, F., {et~al.} 2009, \aap, 506, 377, \dodoi{10.1051/0004-6361/200911897}

\bibitem[{{Triaud} {et~al.}(2013){Triaud}, {Hebb}, {Anderson}, {Cargile}, {Collier Cameron}, {Doyle}, {Faedi}, {Gillon}, {Gomez Maqueo Chew}, {Hellier}, {Jehin}, {Maxted}, {Naef}, {Pepe}, {Pollacco}, {Queloz}, {S{\'e}gransan}, {Smalley}, {Stassun}, {Udry}, \& {West}}]{2013A&A...549A..18T}
{Triaud}, A.~H.~M.~J., {Hebb}, L., {Anderson}, D.~R., {et~al.} 2013, \aap, 549, A18, \dodoi{10.1051/0004-6361/201219643}

\bibitem[{{Tsuribe} \& {Inutsuka}(1999)}]{1999ApJ...523L.155T}
{Tsuribe}, T., \& {Inutsuka}, S.-i. 1999, \apjl, 523, L155, \dodoi{10.1086/312267}

\bibitem[{Umbreit(2005)}]{Umbreit2005The}
Umbreit, S. 2005, \dodoi{10.11588/heidok.00005702}

\bibitem[{{Viani} {et~al.}(2019){Viani}, {Basu}, {Corsaro}, {Ball}, \& {Chaplin}}]{2019ApJ...879...33V}
{Viani}, L.~S., {Basu}, S., {Corsaro}, E., {Ball}, W.~H., \& {Chaplin}, W.~J. 2019, \apj, 879, 33, \dodoi{10.3847/1538-4357/ab232e}

\bibitem[{{{\v{S}}ubjak} {et~al.}(2020){{\v{S}}ubjak}, {Sharma}, {Carmichael}, {Johnson}, {Gonzales}, {Matthews}, {Boffin}, {Brahm}, {Chaturvedi}, {Chakraborty}, {Ciardi}, {Collins}, {Esposito}, {Fridlund}, {Gan}, {Gandolfi}, {Garc{\'\i}a}, {Guenther}, {Hatzes}, {Latham}, {Mathis}, {Mathur}, {Persson}, {Relles}, {Schlieder}, {Barclay}, {Dressing}, {Crossfield}, {Howard}, {Rodler}, {Zhou}, {Quinn}, {Esquerdo}, {Calkins}, {Berlind}, {Stassun}, {Bla{\v{z}}ek}, {Skarka}, {{\v{S}}pokov{\'a}}, {{\v{Z}}{\'a}k}, {Albrecht}, {Sobrino}, {Beck}, {Cabrera}, {Carleo}, {Cochran}, {Csizmadia}, {Dai}, {Deeg}, {de Leon}, {Eigm{\"u}ller}, {Endl}, {Erikson}, {Fukui}, {Georgieva}, {Gonz{\'a}lez-Cuesta}, {Grziwa}, {Hidalgo}, {Hirano}, {Hjorth}, {Knudstrup}, {Korth}, {Lam}, {Livingston}, {Lund}, {Luque}, {Montanes Rodr{\'\i}guez}, {Murgas}, {Narita}, {Nespral}, {Niraula}, {Nowak}, {Pall{\'e}}, {P{\"a}tzold}, {Prieto-Arranz}, {Rauer}, {Redfield}, {Ribas}, {Smith}, {Van Eylen}, \& {Kab{\'a}th}}]{2020AJ....159..151S}
{{\v{S}}ubjak}, J., {Sharma}, R., {Carmichael}, T.~W., {et~al.} 2020, \aj, 159, 151, \dodoi{10.3847/1538-3881/ab7245}

\bibitem[{{Wang} {et~al.}(2022){Wang}, {Rice}, {Wang}, {Pu}, {Stef{\'a}nsson}, {Mahadevan}, {Radzom}, {Giacalone}, {Wu}, {Esposito}, {Dalba}, {Avsar}, {Holden}, {Skiff}, {Polakis}, {Voeller}, {Logsdon}, {Klusmeyer}, {Schweiker}, {Wu}, {Beard}, {Dai}, {Lubin}, {Weiss}, {Bender}, {Blake}, {Dressing}, {Halverson}, {Hearty}, {Howard}, {Huber}, {Isaacson}, {Jackman}, {Llama}, {McElwain}, {Rajagopal}, {Roy}, {Robertson}, {Schwab}, {Shkolnik}, {Wright}, \& {Laughlin}}]{2022ApJ...926L...8W}
{Wang}, X.-Y., {Rice}, M., {Wang}, S., {et~al.} 2022, \apjl, 926, L8, \dodoi{10.3847/2041-8213/ac4f44}

\bibitem[{{Whitworth} \& {Stamatellos}(2006)}]{2006A&A...458..817W}
{Whitworth}, A.~P., \& {Stamatellos}, D. 2006, \aap, 458, 817, \dodoi{10.1051/0004-6361:20065806}

\bibitem[{{Windhorst} {et~al.}(2011){Windhorst}, {Cohen}, {Hathi}, {McCarthy}, {Ryan}, {Yan}, {Baldry}, {Driver}, {Frogel}, {Hill}, {Kelvin}, {Koekemoer}, {Mechtley}, {O'Connell}, {Robotham}, {Rutkowski}, {Seibert}, {Straughn}, {Tuffs}, {Balick}, {Bond}, {Bushouse}, {Calzetti}, {Crockett}, {Disney}, {Dopita}, {Hall}, {Holtzman}, {Kaviraj}, {Kimble}, {MacKenty}, {Mutchler}, {Paresce}, {Saha}, {Silk}, {Trauger}, {Walker}, {Whitmore}, \& {Young}}]{2011ApJS..193...27W}
{Windhorst}, R.~A., {Cohen}, S.~H., {Hathi}, N.~P., {et~al.} 2011, \apjs, 193, 27, \dodoi{10.1088/0067-0049/193/2/27}

\bibitem[{{Winn} {et~al.}(2010){Winn}, {Fabrycky}, {Albrecht}, \& {Johnson}}]{2010ApJ...718L.145W}
{Winn}, J.~N., {Fabrycky}, D., {Albrecht}, S., \& {Johnson}, J.~A. 2010, \apjl, 718, L145, \dodoi{10.1088/2041-8205/718/2/L145}

\bibitem[{{Winn} {et~al.}(2009){Winn}, {Howard}, {Johnson}, {Marcy}, {Gazak}, {Starkey}, {Ford}, {Col{\'o}n}, {Reyes}, {Nortmann}, {Dreizler}, {Odewahn}, {Welsh}, {Kadakia}, {Vanderbei}, {Adams}, {Lockhart}, {Crossfield}, {Valenti}, {Dantowitz}, \& {Carter}}]{2009ApJ...703.2091W}
{Winn}, J.~N., {Howard}, A.~W., {Johnson}, J.~A., {et~al.} 2009, \apj, 703, 2091, \dodoi{10.1088/0004-637X/703/2/2091}

\bibitem[{{Wisdom}(2008)}]{2008Icar..193..637W}
{Wisdom}, J. 2008, \icarus, 193, 637, \dodoi{10.1016/j.icarus.2007.09.002}

\bibitem[{{Wright} {et~al.}(2010){Wright}, {Eisenhardt}, {Mainzer}, {Ressler}, {Cutri}, {Jarrett}, {Kirkpatrick}, {Padgett}, {McMillan}, {Skrutskie}, {Stanford}, {Cohen}, {Walker}, {Mather}, {Leisawitz}, {Gautier}, {McLean}, {Benford}, {Lonsdale}, {Blain}, {Mendez}, {Irace}, {Duval}, {Liu}, {Royer}, {Heinrichsen}, {Howard}, {Shannon}, {Kendall}, {Walsh}, {Larsen}, {Cardon}, {Schick}, {Schwalm}, {Abid}, {Fabinsky}, {Naes}, \& {Tsai}}]{2010AJ....140.1868W}
{Wright}, E.~L., {Eisenhardt}, P.~R.~M., {Mainzer}, A.~K., {et~al.} 2010, \aj, 140, 1868, \dodoi{10.1088/0004-6256/140/6/1868}

\bibitem[{{Wright} {et~al.}(2023){Wright}, {Rice}, {Wang}, {Hixenbaugh}, \& {Wang}}]{2023AJ....166..217W}
{Wright}, J., {Rice}, M., {Wang}, X.-Y., {Hixenbaugh}, K., \& {Wang}, S. 2023, \aj, 166, 217, \dodoi{10.3847/1538-3881/ad0131}

\bibitem[{{Wurster} {et~al.}(2018){Wurster}, {Bate}, \& {Price}}]{2018MNRAS.480.4434W}
{Wurster}, J., {Bate}, M.~R., \& {Price}, D.~J. 2018, \mnras, 480, 4434, \dodoi{10.1093/mnras/sty2212}

\bibitem[{{York} {et~al.}(2000){York}, {Adelman}, {Anderson}, {Anderson}, {Annis}, {Bahcall}, {Bakken}, {Barkhouser}, {Bastian}, {Berman}, {Boroski}, {Bracker}, {Briegel}, {Briggs}, {Brinkmann}, {Brunner}, {Burles}, {Carey}, {Carr}, {Castander}, {Chen}, {Colestock}, {Connolly}, {Crocker}, {Csabai}, {Czarapata}, {Davis}, {Doi}, {Dombeck}, {Eisenstein}, {Ellman}, {Elms}, {Evans}, {Fan}, {Federwitz}, {Fiscelli}, {Friedman}, {Frieman}, {Fukugita}, {Gillespie}, {Gunn}, {Gurbani}, {de Haas}, {Haldeman}, {Harris}, {Hayes}, {Heckman}, {Hennessy}, {Hindsley}, {Holm}, {Holmgren}, {Huang}, {Hull}, {Husby}, {Ichikawa}, {Ichikawa}, {Ivezi{\'c}}, {Kent}, {Kim}, {Kinney}, {Klaene}, {Kleinman}, {Kleinman}, {Knapp}, {Korienek}, {Kron}, {Kunszt}, {Lamb}, {Lee}, {Leger}, {Limmongkol}, {Lindenmeyer}, {Long}, {Loomis}, {Loveday}, {Lucinio}, {Lupton}, {MacKinnon}, {Mannery}, {Mantsch}, {Margon}, {McGehee}, {McKay}, {Meiksin}, {Merelli}, {Monet}, {Munn}, {Narayanan}, {Nash}, {Neilsen}, {Neswold}, {Newberg}, {Nichol}, {Nicinski},
  {Nonino}, {Okada}, {Okamura}, {Ostriker}, {Owen}, {Pauls}, {Peoples}, {Peterson}, {Petravick}, {Pier}, {Pope}, {Pordes}, {Prosapio}, {Rechenmacher}, {Quinn}, {Richards}, {Richmond}, {Rivetta}, {Rockosi}, {Ruthmansdorfer}, {Sandford}, {Schlegel}, {Schneider}, {Sekiguchi}, {Sergey}, {Shimasaku}, {Siegmund}, {Smee}, {Smith}, {Snedden}, {Stone}, {Stoughton}, {Strauss}, {Stubbs}, {SubbaRao}, {Szalay}, {Szapudi}, {Szokoly}, {Thakar}, {Tremonti}, {Tucker}, {Uomoto}, {Vanden Berk}, {Vogeley}, {Waddell}, {Wang}, {Watanabe}, {Weinberg}, {Yanny}, {Yasuda}, \& {SDSS Collaboration}}]{2000AJ....120.1579Y}
{York}, D.~G., {Adelman}, J., {Anderson}, John~E., J., {et~al.} 2000, \aj, 120, 1579, \dodoi{10.1086/301513}

\bibitem[{{Zahn}(1977)}]{1977A&A....57..383Z}
{Zahn}, J.~P. 1977, \aap, 57, 383

\bibitem[{{Zanazzi} {et~al.}(2024){Zanazzi}, {Dewberry}, \& {Chiang}}]{2024arXiv240305616Z}
{Zanazzi}, J.~J., {Dewberry}, J., \& {Chiang}, E. 2024, arXiv e-prints, arXiv:2403.05616, \dodoi{10.48550/arXiv.2403.05616}

\bibitem[{{Zechmeister} \& {K{\"u}rster}(2009)}]{2009A&A...496..577Z}
{Zechmeister}, M., \& {K{\"u}rster}, M. 2009, \aap, 496, 577, \dodoi{10.1051/0004-6361:200811296}

\bibitem[{{Zhou} {et~al.}(2019){Zhou}, {Bakos}, {Bayliss}, {Bento}, {Bhatti}, {Brahm}, {Csubry}, {Espinoza}, {Hartman}, {Henning}, {Jord{\'a}n}, {Mancini}, {Penev}, {Rabus}, {Sarkis}, {Suc}, {de Val-Borro}, {Rodriguez}, {Osip}, {Kedziora-Chudczer}, {Bailey}, {Tinney}, {Durkan}, {L{\'a}z{\'a}r}, {Papp}, \& {S{\'a}ri}}]{2019AJ....157...31Z}
{Zhou}, G., {Bakos}, G.~{\'A}., {Bayliss}, D., {et~al.} 2019, \aj, 157, 31, \dodoi{10.3847/1538-3881/aaf1bb}

\end{thebibliography}
\bibliographystyle{aasjournal}

\end{document}